\pdfoutput=1
\documentclass[12pt]{article}
\usepackage{amsmath,mathrsfs,amsthm}
\usepackage{caption}
\usepackage{verbatim}
\usepackage{subcaption}
\usepackage{graphicx,psfrag,epsf}
\usepackage{enumerate}
\usepackage{natbib}
\usepackage{bbm}
\usepackage{bm}
\usepackage{amssymb}
\usepackage{lscape}
\usepackage{caption}
\usepackage[figuresright]{rotating}
\usepackage[affil-it]{authblk}
\usepackage{verbatim}
\usepackage{extarrows}
\usepackage{framed}
\usepackage[colorlinks=true,citecolor=blue]{hyperref}
\usepackage{algorithm}
\usepackage{algorithmic}
\usepackage{autonum}
\usepackage{url} 
\usepackage{float}
\usepackage{geometry}
\usepackage{lscape} 
\usepackage{multirow}
\usepackage{booktabs}
\usepackage{pifont}
\usepackage{stackengine}
\usepackage{array}
\usepackage{makecell}

\newcommand\pkg[1]{\texttt{#1}}
\newcommand{\blind}{0}
\newcommand{\plainref}[1]{\ref*{#1}}

\newcommand{\R}{\mathbb{R}}
\newcommand{\E}{\mathbb{E}}
\newcommand{\I}{\mathbb{I}}
\newcommand{\Prob}{\mathbb{P}}
\DeclareMathOperator*{\argmax}{arg\,max}
\DeclareMathOperator*{\argmin}{arg\,min}
\newcommand{\ignore}[1]{}
\begin{document}

\def\spacingset#1{\renewcommand{\baselinestretch}%
{#1}\small\normalsize} \spacingset{1}

\date{}
\if0\blind
{
  \title{\bf Simultaneous Estimation of Many Sparse Networks via Hierarchical Poisson Log-Normal Model}
  \author{Changhao Ge\thanks{
    Graduate Group of Applied Mathematics and Computational Science, University of Pennsylvania, \href{mailto:gech@gech.sas.upenn.edu}{gech@gech.sas.upenn.edu}}\hspace{.2cm} and
    Hongzhe Li\thanks{Department of Biostatistics, Epidemiology and Informatics, University of Pennsylvania, \href{mailto:hongzhe@upenn.edu}{hongzhe@upenn.edu}}}
  \maketitle
} \fi

\if1\blind
{
  \bigskip
  \bigskip
  \bigskip
  \begin{center}
    {\LARGE\bf Simultaneous Estimation of Many Sparse Networks via Hierarchical Poisson Log-Normal Model}
\end{center}
  \medskip
} \fi

\bigskip
\begin{abstract}
The advancement of single-cell RNA-sequencing (scRNA-seq) technologies allow us to study the individual level cell-type-specific gene expression networks by direct inference of genes' conditional independence structures. scRNA-seq data facilitates the analysis of gene expression data across different conditions or samples, enabling simultaneous estimation of condition- or sample-specific gene networks. Since the scRNA-seq data are count data with many zeros, existing network inference methods based on Gaussian graphs  cannot be applied to such single cell data directly.  We  propose a hierarchical Poisson Log-Normal model to simultaneously estimate many such networks to effectively incorporate the shared network structures. We develop an efficient simultaneous estimation method that uses the variational EM and alternating direction method of multipliers (ADMM) algorithms, optimized for parallel processing.  Simulation studies show this method outperforms traditional methods in network structure recovery and parameter estimation across various network models. We  apply the method to two single cell RNA-seq datasets, a yeast single-cell gene expression dataset measured under 11 different environmental conditions, and a single-cell gene expression data from 13 inflammatory bowel disease patients. We demonstrate that simultaneous estimation can uncover a wider range of conditional dependence networks among genes, offering deeper insights into gene expression mechanisms. 
\end{abstract}

\noindent%
\vfill

\newpage
\spacingset{1.75} 
\section{Introduction}
New high-throughput single-cell and next-generation sequencing technologies have generated various types of population-level single cell  RNA sequencing (scRNA-seq) data and spatial transcriptomic data. These data provide an integrative understanding of various cell heterogeneity, dynamics and their associations with complex phenotypes.  While many methods have been developed for analysis of scRNA-seq data,  these methods mainly focus on cell-level analysis. For population level scRNA-seq data that are measured over multiple samples or conditions, beyond simple changes in average gene expressions in specific cell types,  one can  estimate sample-specific gene expression distributions and gene expression networks, leading to insights into  a high granularity of changes in expressions, including changes in gene network structures among the genes.  
 Network biology using single-cell data will further accelerate our understanding of cellular heterogeneity \citep{cha2020single}.

scRNA-seq data allow us to estimate the  sample-specific gene  expression  networks using single-cell level expression data of a given sample.   However,  challenge arises due to  limited number of cells for a given  sample, which may result in inaccurate estimations and high variance of the estimated sample-specific gene expression networks. It is therefore important to consider simultaneous  estimation of many sparse gene expression networks. 
The rational is that despite the variability in gene network structures among different samples, these structures often originate from similar biological processes  and thus are expected to share certain underlying structures \citep{witherspoon2007genetic}. Thus, utilizing the information  across all samples/individuals is expected to  enhance the accuracy of estimating gene expression network for each individual. 

Gaussian graphical models have been widely used in quantifying gene expression networks, and estimation of such graphical models has been extensively  studied in recent years. 
\cite{yuan2007ggraphical} introduced a regularization method for sparse precision matrix estimation by minimizing the negative log-likelihood function with an $l_1$-norm penalty on the precision matrix.
\cite{banerjee2008sparse}  developed a fast algorithm using block coordinate descent. The graphical Lasso method \citep{friedman2008glasso} introduced an efficient way to solve this $l_1$ regularization problem iteratively, providing a significant improvement in computational efficiency. 
Other approaches  include  a constrained optimization problem \citep{cai2011clime}, 
 adaptive thresholding \citep{cai2011adaptive} and D-trace loss \citep{zhang2014sparse}.  Rigorous statistical theorems have been developed in  high-dimensional settings \citep{cai2016highd, liu2015highd, meinshausen2006highd}.
 In contrast to frequentist methods, Bayesian approaches address sparsity by employing a sparsity-promoting prior. \cite{gan2019bayesian} and \cite{banerjee2015bayesian} introduced a spike-and-slab prior on the precision matrix, and established a method for sparse estimation.
Beside estimation of Gaussian graphical models,  \cite{allen2013local} proposed a local Poisson graphical model for sequencing count data. In the context of the Poisson Log-Normal model, \cite{chiquet2019variational} introduced a variational approximation method.

Methods for simultaneously estimating multiple Gaussian graphical models have also been developed by imposing  penalty terms to allow some similarities among the  precision matrices.  \cite{patrick2014jointglasso} introduced the fused graphical Lasso and group graphical Lasso. \cite{wang2017fast} contributed a fast algorithm to streamline the underlying computations. Furthermore, \cite{gan2019joint} introduced a Bayesian perspective with hierarchical modeling. However, these works are based on Gaussian graphical models and cannot be directly applied to single-cell gene expression data, which are characterized by sparse count data with many zeros.

To model the gene expression networks based on single cell expression data, we introduce a hierarchical Poisson log-normal (PLN) model, where the gene expression count data is modeled using Poisson distribution with the mean values modeled by a multivariate log-normal distribution with a sparse precision matrix. To leverage the similarity of many  sparse precision matrices in population level single cell studies, we introduce a mixture of Laplacian prior distribution on each element of the precision matrices. 
This allows us to combine likelihoods across all samples \citep{efron1996empirical}, with the goal of improving simultaneous estimation of gene expression networks or gene expression precision matrices for all the samples considered. 

We develop an efficient variational Expectation-Maximization (VEM) algorithm for parameter estimation and obtaining the posterior distribution of the sample-specific precision matrix. The M-steps of the VEM algorithm involve the block-wise coordinate-descent method and the alternating direction method of multipliers (ADMM). 
Our simulation studies indicate that the proposed simultaneous estimation approach improves the accuracy of network structure recovery and achieves higher  Matthews correlation coefficient (MCC) values than methods based  on data from a single sample. We demonstrate our methods by studying the sample-specific gene networks using two single cell RNA-seq datasets. The implementation of our method is available at \href{https://github.com/HowardGech/SimultaneousPLN}{https://github.com/HowardGech/SimultaneousPLN}.

The rest of the paper is organized as follows.  In Section \ref{sec:model}, we introduce the proposed hierarchical Poisson log-normal model. We then describe a VEM algorithm to estimate the parameters and to obtain the posterior estimates of sample-specific precision matrices in Section \ref{sec:computation}. We present the results of simulation studies in Section \ref{sec:simulation} and the findings from the analysis of two  single-cell datasets in Section \ref{sec:real}. Finally, in Section \ref{sec:discussion}, we provide a brief  discussion of the methods and results.

\section{A Hierarchical Poisson Log-Normal Model}\label{sec:model}

\begin{figure}[!t]
    \centering
    \includegraphics[width=0.4\columnwidth]{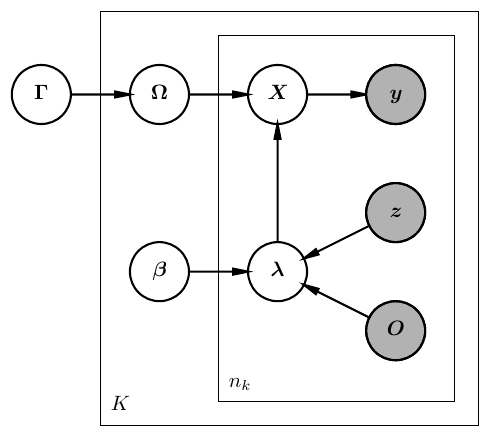}
    \caption{Graphical model representation of the PLN model with Spike-and-Slab precision matrices. The outer plate represents $K$ groups. The inner plate represents $n_k$ single cell observations within each group. Grey nodes denote observed data.}
    \label{fig:graph_structure}
\end{figure}

We first introduce the hierarchical Poisson log-normal (PLN) model for simultaneous  estimation of multiple gene networks for single cell RNA-seq data with a graphical representation provided  in Figure \ref{fig:graph_structure}. 
 Consider $K$ groups of observed data $\{(\bm{y}^{(k)}_i,\bm{z}^{(k)}_i)_{i=1}^{n_k}\}_{k=1}^K$, where  $\bm{y}^{(k)}_i=({y}^{(k)}_{i1},\cdots,{y}^{(k)}_{ip})^\intercal$ $\in\mathbb{N}^{p}$ is the read counts of $p$ genes for the $i$th cell in the $k$th group, $\bm{z}^{(k)}_i=({z}^{(k)}_{i1},\cdots,{z}^{(k)}_{id})^\intercal\in\R^{d}$ represents the $d$-dimensional covariates, and $n_k$ is the number of the cells for the $k$th group. In practice, groups may correspond to categories such as different donors or cell types, while covariates can include clinical information such as age, gender, and race. Additionally, the offset $\bm{o}^{(k)}_{i} =(o^{(k)}_{i1},\cdots,o^{(k)}_{ip})^\intercal\in\R^{p}$ representing  the library sizes is also considered. The data generation follows a two-step process: 
\begin{equation}
\begin{aligned}
    \bm{X}_{i}^{(k)} &\sim N(\bm{\lambda}^{(k)}_{i},(\bm{\Omega}^{(k)})^{-1}), \\
y_{ij}^{(k)} &\sim\mathrm{Pois}(\exp(X_{ij}^{(k)}),
\end{aligned}
\end{equation}
where $\bm{\lambda}_i^{(k)}= \bm{o}_{i}^{(k)}+\bm{\beta}^{(k)}\bm{z}^{(k)}_i$ is the mean vector with coefficient $\bm{\beta}^{(k)}=(\bm{\beta}_1^{(k)},\cdots,\bm{\beta}_p^{(k)})^\intercal\in\R^{p\times d}$, $\bm{\Omega}^{(k)}$ is the precision matrix, and $\mathrm{Pois}(\cdot)$ is the poisson distribution.  We further  assume the precision matrix $\bm{\Omega}^{(k)}$ is sparse for each group $k=1,2,\cdots,K$. The goal is to estimate these precision matrices.

In our examination of multi-sample single-cell data, the sparsity structures of these $K$ precision matrices are similar, though they are not necessarily identical. We propose a Bayesian framework which imposes a prior on all precision matrices to leverage information across all groups for a better estimation. Generally coefficients are not constrained, so their priors are assumed to be non-informative, i.e. $\bm{\beta}^{(k)}\propto1$. To capture the similarity of precision matrices among groups, we follow \cite{gan2019joint} and utilize the idea of latent variables. Define $\gamma_{ij}$ as a binary indicator of whether the $i$-th and $j$-th genes are conditionally dependent $(\gamma_{ij}=1)$ or independent $(\gamma_{ij}=0)$. A Bernoulli distribution with hyperparameter $p_0$ is assumed as the prior for $\gamma_{ij}$. Conditioning on $\gamma_{ij}$, the prior of $\Omega_{ij}^{(k)}$ for each groups $k=1,2,\ldots,K$ follows a Laplace distribution with hyperparameter $v_{\gamma_{ij}}$. In summary, the hierarchical prior for the $(i,j)$-th element ($i\neq j$) of the precision matrix in group $k$ has the form of:
\begin{equation}
\Prob(\Omega_{ij}^{(k)}|\gamma_{ij})=
    \begin{cases}
        &\frac{1}{2v_1}\exp(-{|\Omega_{ij}^{(k)}|}/{v_1}),\quad\text{if }\gamma_{ij}=1,\\
        &\frac{1}{2v_0}\exp(-{|\Omega_{ij}^{(k)}|}/{v_0}),\quad\text{if }\gamma_{ij}=0,
    \end{cases}\qquad\gamma_{ij}\sim \mathrm{Bern}(p_0),
    \label{eq:prior}
\end{equation}
where $v_1>v_0>0$ to  ensure the model is well-defined. For the diagonal elements, we impose an exponential prior with hyperparameter $\tau$ since their values are always positive. 

Denote the matrix $\bm{\Gamma}$ with off-diagonal elements  $\Gamma_{ij}=\gamma_{ij}$ for $i \neq j$, and diagonal elements $\Gamma_{ii}=1$ for all $i=1,2,\ldots,p$. The prior distribution of the precision matrix $\bm{\Omega}^{(k)}$, conditioning on $\bm{\Gamma}$, is given by:
\begin{equation}
    \Prob(\bm{\Omega}^{(k)}|\bm{\Gamma})=\I_{\{\bm{\Omega}^{(k)}\succ0\}}\cdot\prod_{i<j}\left(\mathrm{Lap}(\Omega_{ij}^{(k)};v_{\Gamma_{ij}})\right)\cdot\prod_{i=1}^p\mathrm{Exp}(\Omega_{ii}^{(k)};\tau),
\end{equation}
where $\mathbb{I}_{\{\bm{\Omega}^{(k)}\succ0\}}$ ensures that $\bm{\Omega}^{(k)}$ is positive definite, $\mathrm{Lap}(\cdot;v)$ is the density of Laplace distribution promoting sparsity, and $\mathrm{Exp}(\cdot;\tau)$ is the density of exponential distribution. Conditioning on $\bm{\Gamma}$, the priors for different $\bm{\Omega}^{(k)}$ are independent. The overall prior for the set of precision matrices $\bm{\Omega}:=(\bm{\Omega}^{(k)})_{k=1}^K$ is:
\begin{equation}
    \Prob(\bm{\Omega})=\prod_{k=1}^K\Prob(\bm{\Omega}^{(k)}|\bm{\Gamma})\cdot\prod_{i<j}p_0^{\Gamma_{ij}}(1-p_0)^{1-\Gamma_{ij}}.
\end{equation}

\section{A Variational EM Algorithm for Parameter Estimation}\label{sec:variational-em}

 \subsection{A Variational EM Algorithm}
 A common method for parameter estimation in Bayesian analysis is the Maximum A Posteriori  (MAP) estimate. Given the observed data $\bm{y}:=\{(\bm{y}^{(k)}_i)_{i=1}^{n_k}\}_{k=1}^K$, the posterior distribution of $\bm{\Omega}$ and $\bm{\beta}$ can be derived through integration:
\begin{equation}
    \Prob(\bm{\Omega},\bm{\beta}|\bm{y})\propto\int\prod_{k,i}\left(\prod_{j=1}^p\exp(y^{(k)}_{ij}X^{(k)}_{ij}-\exp(X^{(k)}_{ij}))\cdot N(\bm{X}^{(k)}_i;\bm{\lambda}^{(k)}_{i},(\bm{\Omega}^{(k)})^{-1})d\bm{X}_i^{(k)}\right)\Prob(\bm{\Omega}).
\end{equation}
However, this marginal posterior is typically intractable.

 We propose to approximate the MAP estimator using the EM algorithm.  The details of the following steps are provided in Section \plainref{App.VEM} of the Supplemental Material. Treating $\bm{\Gamma}$ and $\bm{X}:=\{(\bm{X}^{(k)}_i)_{i=1}^{n_k}\}_{k=1}^K$ as hidden variables, the E-step involves constructing a $Q$ function that  is the expectation of log posterior:
\begin{equation}
    \begin{aligned}
    Q(\bm{\Omega},\bm{\beta}|\bm{\Omega}^{(\cdot,t)},\bm{\beta}^{(\cdot,t)})&=\E_{\bm{X},\bm{\Gamma}\sim\Prob(\cdot|\bm{y},\bm{\Omega}^{(\cdot,t)},\bm{\beta}^{(\cdot,t)})}\log\Prob(\bm{\Omega},\bm{\beta}|\bm{y},\bm{X},\bm{\Gamma})\\
    &=\E_{\bm{X},\bm{\Gamma}\sim\Prob(\cdot|\bm{y},\bm{\Omega}^{(\cdot,t)},\bm{\beta}^{(\cdot,t)})}\log\Prob(\bm{y},\bm{X}|\bm{\Omega},\bm{\beta})\Prob({\bm{\Omega}|\bm{\Gamma}})+C,
    \end{aligned}
\end{equation}
    where $C$ is a constant irrelevant to $\bm{\Omega}$, and the log complete data likelihood function is
\begin{equation}
\log\Prob(\bm{y},\bm{X}|\bm{\Omega},\bm{\beta})=\sum_{k=1}^K\log\Prob(\bm{y}^{(k)}|\bm{X}^{(k)})\Prob(\bm{X}^{(k)}|\bm{\Omega}^{(k)},\bm{\beta}^{(k)}).
\label{eq:elbo}
    \end{equation}
    Therefore, the $Q$ function  can be derived  as (up to a constant):
\begin{equation}
\begin{aligned}
Q(\bm{\Omega},\bm{\beta}|\bm{\Omega}^{(\cdot,t)},\bm{\beta}^{(\cdot,t)})&=\sum_{k=1}^K\E_{\bm{X}^{(k)}\sim\Prob(\cdot|\bm{y}^{(k)},\bm{\Omega}^{(k,t)},\bm{\beta}^{(k,t)})}\log\Prob(\bm{X}^{(k)}|\bm{\Omega}^{(k)},\bm{\beta}^{(k)})\\
&+\E_{\bm{\Gamma}\sim\Prob(\cdot|\bm{\Omega}^{(\cdot,t)})}\log\Prob({\bm{\Omega}|\bm{\Gamma}}).
    \label{eq:Qfunction}
\end{aligned}
\end{equation}
However, this $Q$ function lacks an explicit form. In fact, denoting the density of Poisson distribution with mean $\lambda$ as $\mathrm{Pois}(\cdot;\lambda)$, the normalizing constant has the expression of
\begin{equation}
    \Prob(\bm{y}_{i}^{(k)}|\bm{\Omega}^{(k,t)},\bm{\beta}^{(k,t)})=\int\prod_{j=1}^p\mathrm{Pois}(y_{ij}^{(k)};\exp(-e^{X_{ij}^{(k)}}))\cdot N(\bm{X}_{i}^{(k)};\bm{\lambda}_i^{(k,t)},(\bm{\Omega}^{(k,t)})^{-1})d\bm{X}_{i}^{(k)},
\end{equation}
which includes an Exponential integral (Ei) that is not an elementary function, resulting in an intractable conditional probability $\Prob(\bm{X}^{(k)}|\bm{y}^{(k)},\bm{\Omega}^{(k,t)},\bm{\beta}^{(k,t)})$. 

To address this challenge, we adopt a variational approximation of $\bm{X}$. Consider the following family of distributions:
\begin{equation}
    \mathcal{P}_{VI}=\left\{P(\bm{X})=\prod_{k=1}^K\prod_{i=1}^{n_k}\prod_{j=1}^pN({{X}^{(k)}_{ij}};\mu_{ij}^{(k)},\sigma_{ij}^{2(k)})\right\},
    \label{eq:variational_family}
\end{equation}
where $N(\cdot;\mu,\sigma^2)$ represents the density of normal distribution with mean $\mu$ and variance $\sigma^2$. The variational posterior of $\bm{X}$ is defined as
\begin{equation}
   \widetilde{\Pi}^{(\cdot,t)} := \argmin_{P\in\mathcal{P}_{VI}}\quad D_{\text{KL}}(P(\bm{X})\|\Prob(\bm{X}|\bm{y},\bm{\Omega}^{(\cdot,t)},\bm{\beta}^{(\cdot,t)})),
   \label{eq:tilde_pi}
\end{equation}
where  $D_{\text{KL}}(\cdot\|\cdot)$ is the KL divergence. Replacing the posterior $\Prob(\bm{X}^{(k)}|\bm{y}^{(k)},\bm{\Omega}^{(k,t)},\bm{\beta}^{(k,t)})$ in (\ref{eq:Qfunction}) by the variational surrogate $\widetilde{\Pi}^{(k,t)}$, an approximation of $Q$ function is expressed as:
\begin{equation}
    \widetilde{Q}(\bm{\Omega},\bm{\beta}|\bm{\Omega}^{(\cdot,t)},\bm{\beta}^{(\cdot,t)})=\sum_{k=1}^K\E_{\bm{X}^{(k)}\sim\widetilde{\Pi}^{(k,t)}}\log\Prob(\bm{X}^{(k)}|\bm{\Omega}^{(k)},\bm{\beta}^{(k)})+\E_{\bm{\Gamma}\sim\Prob(\cdot|\bm{\Omega}^{(\cdot,t)})}\log\Prob({\bm{\Omega}|\bm{\Gamma}}),
    \label{eq:Qtilde}
\end{equation}
where
\begin{equation}
    \log\Prob(\bm{X}^{(k)}|\bm{\Omega}^{(k)},\bm{\beta}^{(k)})=\frac{n_k}{2}\log|\bm{\Omega}^{(k)}|-\frac{1}{2}\sum_{i=1}^{n_k}(\bm{X}_{i}^{(k)}-\bm{\lambda}_{i}^{(k)})^\intercal\bm{\Omega}^{(k)}(\bm{X}_i^{(k)}-\bm{\lambda}_i^{(k)}).
\end{equation}
 The M-step of the above variational EM algorithm takes $(\bm{\Omega}^{(\cdot,t+1)},\bm{\beta}^{(\cdot,t+1)})=\argmax_{(\bm{\Omega},\bm{\beta})}\widetilde{Q}(\bm{\Omega},\bm{\beta}|\bm{\Omega}^{(\cdot,t)},\bm{\beta}^{(\cdot,t)})$. This process is repeated until convergence.

\subsection{Hyperparameter Selection}\label{sec:tuning-param}

Our model involves three hyperparameters: $p_0$, governing the overall sparsity of the networks, and $v_0$ and $v_1$, regulating the penalty of each element in the precision matrices. An improper choice of these hyperparameters could impair the performance of our method. Typically, these  parameters are determined through cross-validation (CV), where the hyperparameters yielding the smallest CV error are suggested for practical use. However, CV requires a predictive error metric or a reliable approximation, which is not feasible in hierarchical models like the PLN model due to their inherent unknown hidden states. Furthermore, for the PLN model, integrating over these hidden states to compute such an error metric is not possible because the required integral is mathematically intractable.

An alternative strategy involves employing model selection criteria such as the Akaike's information criterion (AIC, \cite{Akaike1992aic}), and the Bayesian information criterion (BIC, \cite{Gideon1978bic}), which balance model fit and the number of parameters. The extended Bayesian information criterion (EBIC, \cite{chen2008ebic}) expanded this approach by also considering model space complexity, offering reliable performance in variable selection within large model spaces. For our model, the EBIC for the $k$th group is defined as follows:
\begin{equation}
    \text{EBIC}^k_{\gamma}(\bm{\beta}^{(k)},\bm{\Omega}^{(k)})=-2 l^{(k)}(\bm{\beta}^{(k)},\bm{\Omega}^{(k)})+(|E^{(k)}|+pd)\log n_k+\gamma\log{p(p+1)/2 \choose |E^{(k)}|},
\end{equation}
where $l^{(k)}=\sum_{i=1}^{n_k}\log\Prob(\bm{y}^{(k)}_i|\bm{\beta}^{(k)},\bm{\Omega}^{(k)})$ is the log likelihood, $E^{(k)}$ is the edge set of the estimated network, and $\gamma\in[0,1]$ controls the penalty of graph sparsity. In our implementation we set $\gamma=0.5$.

While model selection criteria like AIC, BIC, and EBIC are fundamental statistical tools, their application to our model is not straightforward because  the full log-likelihood function is not computationally feasible. Following the approach of \cite{chiquet2019variational}, we address this issue by replacing the log-likelihood term $l^{(k)}$ by its variational approximation $\E_{\bm{X}_i^{(k)}\sim\widetilde{\Pi}_i^{(k,t)}}\log\Prob(\bm{y}_i^{(k)}|\bm{X}_i^{(k)})\Prob(\bm{X}_i^{(k)}|\bm{\Omega}^{(k)},\bm{\beta}^{(k)})$. In practice, we conduct a grid search to identify the hyperparameter set that minimizes the (variational) EBIC value.

\section{Computational Implementation}\label{sec:computation}

This section provides a detailed derivation of the optimization algorithm using coordinate ascent. In each iteration, the ADMM method is applied to compute variational mean vectors. We show that this algorithm is parallelizable, making it efficient on multi-core computing systems. The detailed algorithms as well as their derivations are given in Supplementary Section \plainref{supp:sec:algo_and_details}.

\subsection{Maximizing the Variational Function}
The variational $Q$ function \eqref{eq:Qtilde} comprises two components: the likelihood term, which involves computing the variational posterior \eqref{eq:tilde_pi}, and the prior term. The latter one requires knowing the conditional distribution $\Prob(\bm{\Gamma}|\bm{\Omega}^{(\cdot,t)})$, which can be calculated as:
\begin{equation}
\label{eq:posterior_gamma}
    \Prob(\gamma_{ij}=1|\bm{\Omega}^{(\cdot,t)})=1/\left(1+\frac{1-p_0}{p_0}(\frac{v_1}{v_0})^K\exp\left(-(\frac{1}{v_0}-\frac{1}{v_1})\sum_{k=1}^K|\Omega_{ij}^{(k,t)}|\right)\right).
\end{equation}
Denoting this probability as $p_{ij}(\bm{\Omega}^{(\cdot,t)})$, the prior term is given by:
\begin{equation}
\E_{\bm{\Gamma}\sim\Prob(\cdot|\bm{\Omega}^{(\cdot,t)})}\log\Prob({\bm{\Omega}|\bm{\Gamma}})=\sum_{i<j}\left(\frac{p_{ij}(\bm{\Omega}^{(\cdot,t)})}{v_1}+\frac{1-p_{ij}(\bm{\Omega}^{(\cdot,t)})}{v_0}\right)|\Omega_{ij}|+\sum_{i=1}^p\frac{1}{\tau}\Omega_{ii}.
\end{equation}

To compute the likelihood term in \eqref{eq:Qtilde}, it is necessary to determine the variational posterior for $\bm{X}$. Defining $\bm{\mu}_i^{(k)}=(\mu_{i1}^{(k)},\mu_{i2}^{(k)},\cdots,\mu_{ip}^{(k)})^\intercal$ and $\bm{\sigma}_i^{2(k)}=(\sigma_{i1}^{2(k)},\sigma_{i2}^{2(k)},\cdots,\sigma_{ip}^{2(k)})^\intercal$, the target function in \eqref{eq:tilde_pi} can be broken down into a sum:
\begin{equation}
    D_{\text{KL}}(P(\bm{X})\|\Prob(\bm{X}|\bm{y},\bm{\Omega}^{(\cdot,t)},\bm{\beta}^{(\cdot,t)}))=\sum_{k=1}^K\sum_{i=1}^{n_k}D_{\text{KL}}(P(\bm{X}^{(k)}_i)\|\Prob(\bm{X}^{(k)}_i|\bm{y}^{(k)}_i,\bm{\Omega}^{(k,t)},\bm{\beta}^{(k,t)})),
    \label{eq:full-KL}
\end{equation}
where
\begin{equation}
    \begin{aligned}
    D_{\text{KL}}(P(\bm{X}_i^{(k)})\|\Prob(\bm{X}_i^{(k)}|\bm{y}_i^{(k)},\bm{\Omega}^{(k,t)},\bm{\beta}^{(k,t)}))&=\frac{1}{2}(\bm{\mu}_i^{(k)}-\bm{\lambda}^{(k,t)}_{i})^\intercal\bm{\Omega}^{(k,t)}(\bm{\mu}_i^{(k)}-\bm{\lambda}^{(k,t)}_{i})-\bm{y}^{(k)\intercal}_i\bm{\mu}_i^{(k)}\\
    &+\frac{1}{2}\sum_{j=1}^p\left(\sigma_{ij}^{2(k)}\Omega_{jj}^{(k,t)}+2\exp(\mu_{ij}^{(k)}+\frac{\sigma^{2(k)}_{ij}}{2})-\log\sigma_{ij}^{2(k)}\right).
    \label{eq:KL_detail}
    \end{aligned}
\end{equation}
It is challenging to optimize the $\bm{\mu}_i^{(k)}$ and $\bm{\sigma}^{2(k)}$ simultaneously because they are entangled in the objective function. To address this issue, we first update the variational mean $\bm{\mu}_i^{(k)}$ while keeping $\bm{\sigma}^{2(k)}$ fixed, and then calculate the variational variance $\bm{\sigma}^{2(k)}$ using the updated $\bm{\mu}_i^{(k)}$.

After obtaining the variational posterior and corresponding parameters
\begin{equation}
    ({\bm{\mu}}^{(k,t)},{\bm{\sigma}}^{(k,t)})=\argmin_{{\bm{\mu}}^{(k)},{\bm{\sigma}}^{(k)}}D_{\text{KL}}(P(\bm{X}_i^{(k)})\|\Prob(\bm{X}^{(k)}|\bm{y}^{(k)},\bm{\Omega}^{(k,t)},\bm{\beta}^{(k,t)})),
\end{equation}
optimizing $\Tilde{Q}(\bm{\Omega},\bm{\beta}|\bm{\Omega}^{(\cdot,t)},\bm{\beta}^{(\cdot,t)})$ can be separated into two parts. First, The update rule of $\bm{\beta}$ is given by
\begin{equation}
    {\bm{\beta}}^{(k,t+1)} = \left({\bm{\mu}}^{(k,t)}-\bm{o}^{(k)}\right)\bm{z}^{(k)\intercal}\left(\bm{z}^{(k)}\bm{z}^{(k)\intercal}\right)^{-1},
\end{equation}
which is irrelevant to the precision matrix. Updating $\bm{\Omega}^{(k)}$ thus becomes a weighted graphical Lasso problem, which can be solved using either the least angle regression technique \citep{friedman2008glasso} or the quadratic approximation method \citep{hsieh2014quic}. The full algorithm is provided in Supplementary Algorithm \plainref{alg:Omega}.

\subsection{ADMM Update for the Variational Mean}\label{sec:admm}
It remains to get an update rule for variational parameters to complete the variational EM algorithm. As shown in \eqref{eq:KL_detail}, the estimation of $\bm{\sigma}_i^{2(k)}$ can be decomposed into $p$ univariate optimization problems and can be solved by efficient algorithms. However, updating the variational mean $\bm{\mu}_i^{(k,t)}$ is not straightforward, which doesn't have an explicit solution and requires multivariate optimization. One potential solution is to use the element-wise coordinate descent algorithm. However, this approach can be computationally expensive, particularly when the dimension $p$ is high. We instead adopt the ADMM algorithm \citep{boyd2011ADMM}, which was  applied in related works on graphical model estimation \citep{patrick2014jointglasso,mohan2012structuredglasso,tang2022singlecell}. 

The optimization problem for the variational mean vector $\bm{\mu}_i^{(k,t)}$ can be reformulated as
\begin{equation}
\begin{aligned}
    \min_{\bm{\mu}_N,\bm{\mu}_M} \ell_1(\bm{\mu}_N,\bm{\Omega}^{(k,t)},\bm{\lambda}^{(k,t)}_i)&+\sum_{j=1}^{p}\ell_2(\mu_{Mj},y_{ij}^{(k)},\sigma_{ij}^{(k,t-1)})\\
   \text{subject to } &\bm{\mu}_N = \bm{\mu}_M
\end{aligned}
\label{eq:admm}
\end{equation}
where
\begin{equation}
    \ell_1(\bm{\mu},\bm{\Omega},\bm{\lambda}) =\frac{1}{2}(\bm{\mu}-\bm{\lambda})^\intercal\bm{\Omega}(\bm{\mu}-\bm{\lambda}),
\end{equation}
and
\begin{equation}
    \ell_2(\mu,y,\sigma^2)=-\mu y+\exp\left(\mu+\frac{\sigma^2}{2}\right).
\end{equation}
Denoting the objective function in \eqref{eq:admm} as $\ell^{(k)}(\bm{\mu}_N,\bm{\mu}_M)$, the ADMM algorithm aims at finding its minimum by optimizing the augmented Lagrangian
\begin{equation}
    \ell^{(k)}(\bm{\mu}_N,\bm{\mu}_M)+\bm{\alpha}^\intercal\left(\bm{\mu}_N-\bm{\mu}_M\right)+\frac{\rho}{2}\|\bm{\mu}_N-\bm{\mu}_M\|^2,
\end{equation}
 with respect to $\bm{\mu}_N$, $\bm{\mu}_M$ and $\bm{\alpha}$ in a coordinate-wise manner. Here $\bm{\alpha}$ is the Lagrangian multiplier and $\rho$ is the step size. This process is terminated when the difference of $\bm{\mu}_N$ and $\bm{\mu}_M$ is below a prespecified threshold. The ADMM algorithm is detailed in Supplementary Algorithm \plainref{alg:ADMM}.
 
 \ignore{The convergence  of ADMM algorithm for this problem is given by the following theorem:
 \begin{theorem}
    For any $\bm{\lambda}^{(k)}_i\in\R^p, \bm{y}_i^{(k)}\in\mathbb{N}^p$, $ \bm{\sigma}_i^{(k)}\in\R^p_{+}$ and positive definite $\bm{\Omega}^{(k)}\in\R^{p\times p}$, the optimization problem \eqref{eq:admm} has a unique solution $\bm{\mu}_M = \bm{\mu}_N = \bm{\mu}^\star\in\R^p$, and this optimal value is attained by Algorithm \ref{alg:ADMM} if the threshold $\epsilon\rightarrow 0$ and maximum iteration number $T\rightarrow\infty$. Furthermore,
    \begin{equation}
        \begin{gathered}
            |\ell^{(k)}(\bar{\bm{\mu}}_N^{(t)},\bar{\bm{\mu}}_M^{(t)})-\ell^{(k)}(\bm{\mu}^\star,\bm{\mu}^\star)|=O(1/t),\\
            |\bar{\bm{\mu}}_N^{(t)}-\bar{\bm{\mu}}_M^{(t)}|=O(1/t),
        \end{gathered}
    \end{equation}
    where $\bar{\bm{\mu}}_N^{(t)}$ and $\bar{\bm{\mu}}_M^{(t)}$ are defined in Algorithm \ref{alg:ADMM}.
    \label{thm:ADMM-convergence}
\end{theorem}
}

\ignore{Theorem \ref{thm:ADMM-convergence} provides the $O(1/t)$ convergence rate for the ADMM algorithm in our method. This result is based on the Theorem 15.4 in \cite{Amir2017FirstOrder} and the fact that our loss function $\ell_1$ and $\ell_2$ are closed and convex. The detailed derivation of this theorem can be found in the Supplementary Material.
}

\subsection{Parallel Computing}

Our coordinate descent  algorithm is summarized as follows. First, we update the penalty matrix $\bm{P}$ utilizing information from all precision matrices. Then, for each sample and every observation within, we update the variational mean and variance. The last step is a weighted graphical Lasso for each sample. These steps are repeated until convergence.

One advantage of our algorithm is that it can take the advantage of parallel computing. This is particularly beneficial when dealing with a large number of groups  $K$, sample size $n_k$, or dimensionality $p$. To elucidate, notice that for each sample, the update of ${\mu}_{Mj}$ within the ADMM step, for different values of $j$, does not interfere with each other. Hence, these updates can be executed simultaneously. This same rationale applies to the update of ${\sigma}_{i,j}^2$. Furthermore, within each sample, given the current estimation of the precision matrix, the variational mean and variance of different observations are independent of each other. Moreover, the updates of $K$ precision matrices are also decoupled, except for the computation of penalty matrix $\bm{P}$. These structural characteristics imply that this algorithm can be highly parallelized.

\section{Simulation Studies}\label{sec:simulation}
We evaluate the performance of our algorithm under three simulation settings. 
For each  setting, the hyperparameters for our method are set as follows: the inclusion probability $p_0$ is fixed at 0.5, and the parameter $\tau$ for the diagonal term is set to infinity, implying no penalty for the diagonal elements. For the off-diagonal penalty parameters, we maintain a fixed ratio of $v_1/v_0=10$, and consider a range of $v_0\in\{0.1, 0.25, 0.5, 1, 5\}\times\sqrt{K\log p/(\sum_{k=1}^Kn_k)}$. The hyperparameter $v_0$ is selected by the EBIC discussed in Section \ref{sec:tuning-param}. An additional evaluation of parallel computing is reported in Supplementary Figure \plainref{fig:computation}.

In our assessment, we use TP, TN, FP, and FN as abbreviations for true positives, true negatives, false positives, and false negatives, respectively, in the context of variable selection. We define $\textnormal{precision}=\textnormal{TP}/(\textnormal{TP}+\textnormal{FP})$ and $\textnormal{recall}=\textnormal{TP}/(\textnormal{TP}+\textnormal{FN})$, to measure the model's accuracy. To evaluate the model's overall performance, we also utilize MCC, which is defined by:
\begin{equation}
\text{MCC} = \frac{\text{TP} \times \text{TN} - \text{FP} \times \text{FN}}{\sqrt{(\text{TP} + \text{FP})(\text{TP} + \text{FN})(\text{TN} + \text{FP})(\text{TN} + \text{FN})}}.
\label{eq:MCC}
\end{equation}
In cases where the true model is the PLN model, we also report the matrix estimation error of the off-diagonal elements, measured by the Frobenius norm (MOFE). All results are based on 200 independent runs.

\subsection{Erdős–Rényi Graph}\label{sec:simulation_pln}

\begin{figure}[!t]
    \centering
    \includegraphics[width=0.98\textwidth]{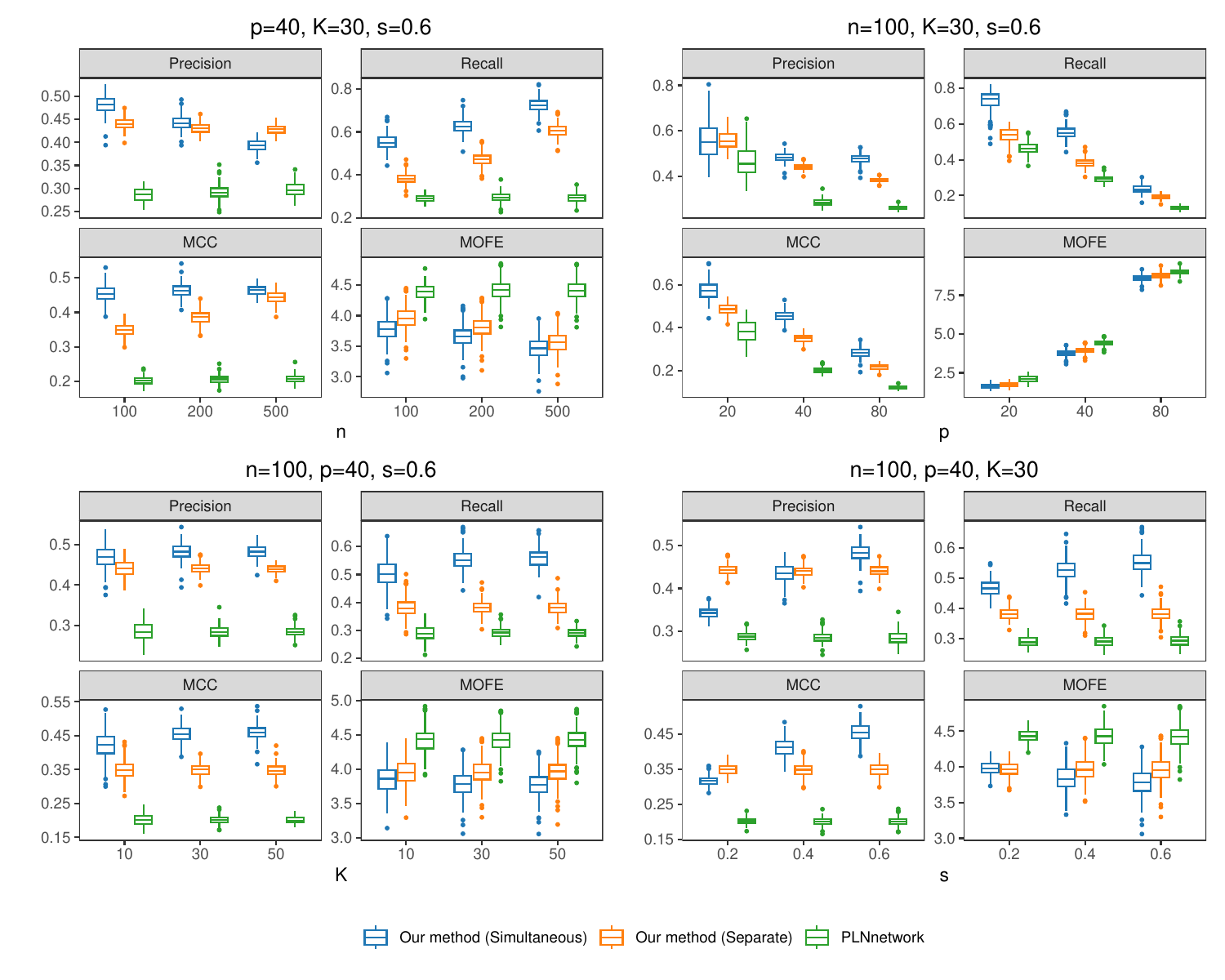}
    \caption{Simulation results under the Erdős–Rényi graph with various values of $n,p,K$ and $s$. For each setting, the result is based on 200 replicates.}
    \label{fig:comp_boxplot}
\end{figure}

We examine the performance of our method when the underlying data-generating model is the PLN model. Our primary focus is on comparing our method, which includes both simultaneous estimation and separate estimation scenarios (i.e., $K=1$), and the PLNnetwork algorithm \citep{chiquet2019variational}, implemented in the R package \pkg{PLNmodels}. The hyperparameter for PLNnetwork is also chosen to minimize the EBIC. Both methods are based on the variational EM framework. However, their penalty structures are different. While our approach allows for element-specific sparsity patterns, PLNnetwork imposes an isotropic penalty on all off-diagonal elements, limiting its ability to capture similar sparsity patterns across multiple samples.

 We construct a model with a simultaneous sparse pattern, which is a key assumption of our algorithm. The true networks in this scenario are generated as follows. We first create an Erdős–Rényi graph $G$, with dimension $p$ and inclusion probability $0.1/s$. This graph represents an underlying structure shared by all samples. Additionally, we generate $K$ independent $p$-dimensional Erdős–Rényi graphs $\{G_k\}_{k=1}^K$, each with an inclusion probability $s$, where $s\in(0.1,1)$ controls the similarity among different samples.
For each group $k=1,2,\cdots,K$, the graph adjacency matrix is the element-wise product of the adjacency matrices of $G$ and $G_k$. The values of nonzero elements are generated randomly, as described in Section \plainref{App.simu} of the Supplementary Material. Each group contains $n$ independent and identically distributed ($i.i.d.$) observations drawn from the PLN model.

We conduct a comprehensive evaluation under varying conditions with $n\in\{100,200,500\}$, $p\in\{20,40,80\}$, $K\in\{10,30,50\}$, and $s\in\{0.2,0.4,0.6\}$. The results are summarized in Figure \ref{fig:comp_boxplot}. An individual-level comparison of simultaneous estimation and separate estimation of our method is reported in the  Figure \plainref{fig:comp_individual}.
Figure \ref{fig:comp_boxplot} shows that simultaneous estimation consistently outperforms alternative approaches across most scenarios, demonstrating its effectiveness even with a relatively small number of samples ($K$). This indicates the advantage of simultaneous estimation when precision matrices exhibit similarities. Furthermore, we observe that simultaneous estimation  improves the estimate of sample-specific precision matrices.  However, one exception occurs for the case $n=100, p=40, K=30, s=0.2$, where separate estimation surpasses simultaneous estimation in MCC and precision. This suggests that simultaneous modeling may not always be beneficial, particularly when the tasks lack a common structure.


\subsection{More Graph Structures}\label{sec:simulation_pln_various}
We next consider different random graph structures and fix values of $n=100, p=40, K=30$. 

\begin{enumerate}[ {(}a{)} ]
\itemsep0em 
    \item  Blocked Graph: Inspired by \cite{tang2022singlecell}, we divide all nodes of a graph into 5 blocks of equal sizes. For each group, nodes in the same block share an edge with probability 0.8. Nodes that reside in different blocks do not have connections between them.
    \item  Hub Graph: In this configuration, $10\%$ of the nodes are designated as hub nodes. For each group, hub node is connected to another node with probability 0.8, while non-hub nodes are not connected to each other.
    \item Scale-free Graph: We first create a random graph using the Barabasi-Albert model \citep{barabasi1999emergence} with parameter $p/8$ to establish an underlying structure. Then, for each group, we independently generate a graph with parameter $p/2$. The adjacency matrix for each graph is the element-wise product of the common matrix and the group-specific matrix.
    \item Small-world Graph: The Watts–Strogatz model \citep{watts1998collective} is known for its small-world property, characterized by a high clustering coefficient and low distances. We create the baseline graph using the Watts-Strogatz model with mean degree $p/5$ and rewiring probability 0.5. For each group, we draw another graph with mean degree $p/4$ and rewiring probability 0.5. The product of two graphs is set as the true graph.
\end{enumerate}

\begin{figure}[!t]
    \centering
    \includegraphics[width=0.98\textwidth]{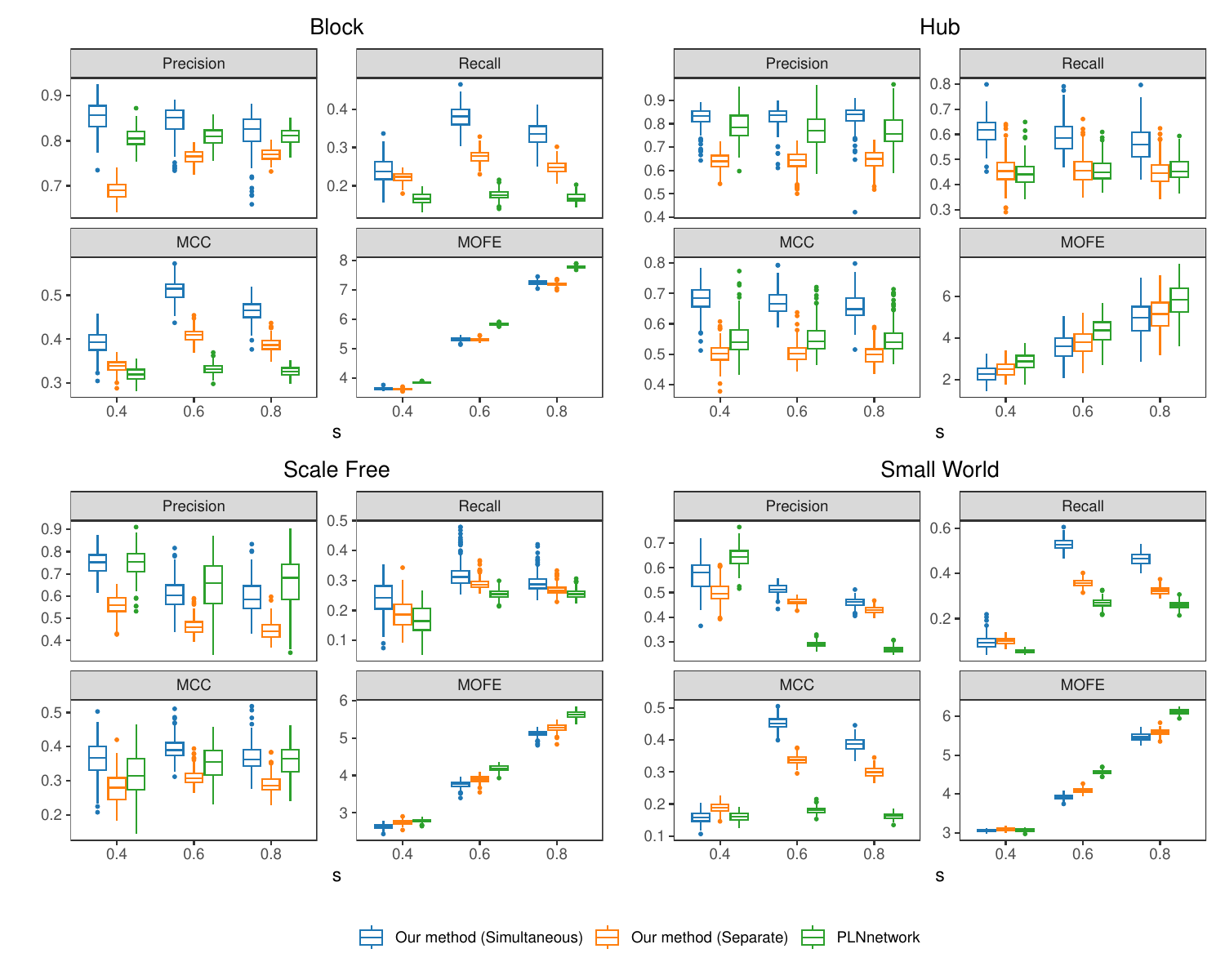}
    \caption{Simulation results under 4 different graph structures, with $n=100, p=40, K=30$ and $s\in\{0.4, 0.6, 0.8\}$.}
    \label{fig:various_structures}
\end{figure}

In each setting, the diagonal elements of the precision matrix are randomly generated from a uniform distribution $\mathrm{Unif}[1,1.5]$. For the nonzero off-diagonal elements, we first sample a value uniformly in $[-0.8, -0.3] \cup [0.3, 0.8]$, then multiply this value by $s$. The parameter $s \in (0, 1)$ represents the signal strength. Positive definiteness is ensured by adding another diagonal matrix (see Section \plainref{App.simu} of the Supplementary Material).

The simulation results  are summarized in Figure \ref{fig:various_structures}. 
Our method outperforms  the PLNnetwork for  all four evaluation criteria, whether estimated simultaneously or individually. This demonstrates our method's ability to capture not only the common structure among all samples but also each individual graph accurately.
 Its greatest gain is achieved in Hub Graphs, a structure common in gene expression networks, social networks, and various other applications, where it excels under all four evaluation criteria. However, in scenarios of small world graphs with low signal strength, simultaneous estimation might have low recall and MCC compared to individual estimation. This suggests that the benefits of simultaneous modeling may diminish with very weak signals.

\subsection{Model Misspecification}\label{sec:misspec}

To test our method's robustness to model misspecification, we simulate count data from another hierarchical model. This involves:
\begin{enumerate}
\itemsep0em 
    \item Generating precision matrices and coefficients as outlined in Section \ref{sec:simulation_pln}, and sampling latent variables $\bm{X}_{i}^{(k)}$ per Section \ref{sec:model}.
    \item Computing $q_{ij}^{(k)}=\exp(X_{ij}^{(k)})/\sum_{l=1}^p\exp(X_{il}^{(k)})$, and defining $\bm{q}_i^{(k)}:=(q_{i1}^{(k)},q_{i2}^{(k)},\cdots,q_{ip}^{(k)})$.
    \item For each group $k$, drawing $\bm{y}_{i}^{(k)}$ independently from a multinomial distribution with size $2p$ and probability $\bm{q}_i^{(k)}$.
\end{enumerate}
We set $n=100$, $p=40$, $K=30$ and $s=0.6$ for our simulation, applying both our method and PLNnetwork. Additionally, the Graphical Lasso (GLasso), Joint Graphical Lasso (JGLasso), and Poisson Graphical Lasso (PGLasso) from the \pkg{glasso}, \pkg{JGL}, and \pkg{RNAseqNet} packages are evaluated. Hyperparameters for GLasso and JGLasso are selected based on the lowest EBIC in the Gaussian graphical model, while PGLasso's hyperparameter follows the stability approach to regularization selection \citep{liu2010stability}. The results are summarized in Table \ref{tab:misspec}.

\begin{table}[!t]
\centering
\caption{The simulation result of MCC, precision and recall under model misspecification, with standard errors in parentheses. Results are based on 200 replicates.}
\begin{tabular}{lccc}
\toprule
  & MCC & Precision & Recall\\
\midrule
Our method (Simultaneous) & 0.264(0.018) & 0.246(0.023) & 0.570(0.047)\\
Our method (Separate) & 0.209(0.014) & 0.258(0.015) & 0.350(0.018)\\
PLNnetwork & 0.195(0.010) & 0.262(0.013) & 0.313(0.019)\\
GLasso & 0.147(0.006) & 0.150(0.010) & 0.659(0.017)\\
JGLasso & 0.140(0.007) & 0.148(0.010) & 0.637(0.016)\\
PGLasso & 0.235(0.015) & 0.573(0.026) & 0.136(0.015)\\
\bottomrule
\end{tabular}
\label{tab:misspec}
\end{table}

Among all the methods considered, our simultaneous estimation approach achieves the highest MCC, with PGLasso as a close second. PLNnetwork shows lower MCC scores, while GLasso and JGLasso lag significantly behind. Although PGLasso demonstrates high precision, it suffers from low recall, indicating fewer edge selections, which may lead to many false negatives in identifying related genes. These results underscore the robustness of our method, particularly in cases where the underlying model is misspecified—a common challenge in practical applications. Interestingly, JGLasso performs slightly worse than GLasso in this scenario, despite the presence of shared group patterns. This suggests that joint modeling in graphical lasso may amplify misspecification errors, in contrast to our approach.

\ignore{
\begin{figure}[!t]
    \centering
    \includegraphics[width=0.7\linewidth]{figure/degree_hist.pdf}
    \caption{Estimated degree distributions for the yeast gene expression networks by PLNnetwork and simultaneous estimation.}
    \label{fig:degree}
    \end{figure}
  }

\section{Analysis of Sample-specific Gene Expression Networks Based on scRNA-seq Data}\label{sec:real}

 We conduct gene network analyses on two single-cell RNA-seq datasets. First, we examine yeast single-cell data collected under various in vitro environmental growth conditions with different nitrogen and carbon sources. Second, we investigate human gene networks in T-cells from patients with ulcerative colitis, comparing the networks of inflamed and non-inflamed cells. For both studies, we present our findings using the simultaneous estimation method and the PLNnetwork, demonstrating the advantage of simultaneous estimation in capturing biological processes. Supplementary Section \plainref{supp:sec:supporting} provides supporting data for these studies.

\subsection{Yeast Single Cell Data Analysis}

\begin{figure}[!t]
    \centering
    \begin{subfigure}[b]{.9\linewidth}
        \includegraphics[width=\linewidth]{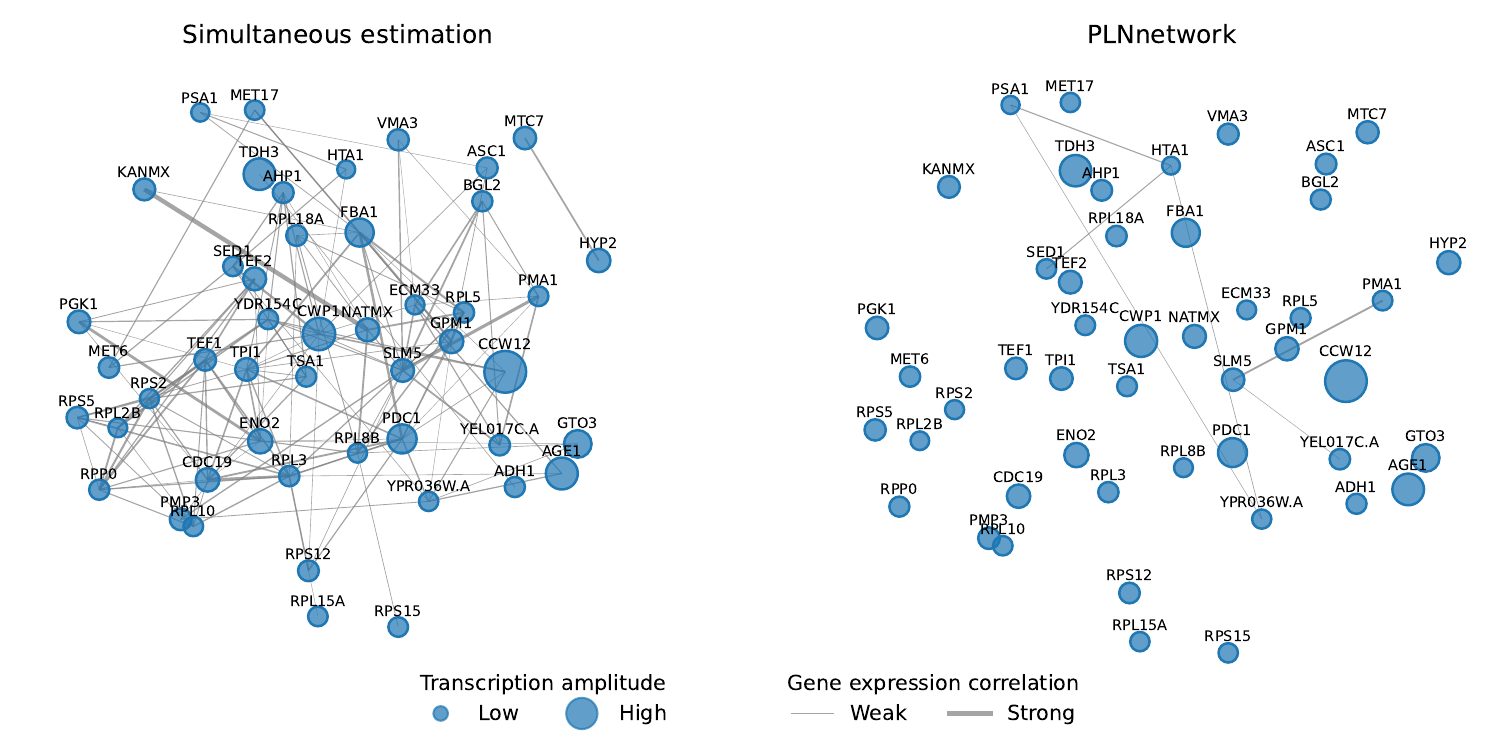}

    \caption{Yeast gene expression network for mostly expressed genes under AmmoniumSulfate condition. Left is estimated by our method. Right is estimated by PLNnetwork.    \label{fig:graph_main}}
    \end{subfigure}
    \begin{subfigure}[b]{.9\linewidth}
            \includegraphics[width=\linewidth]{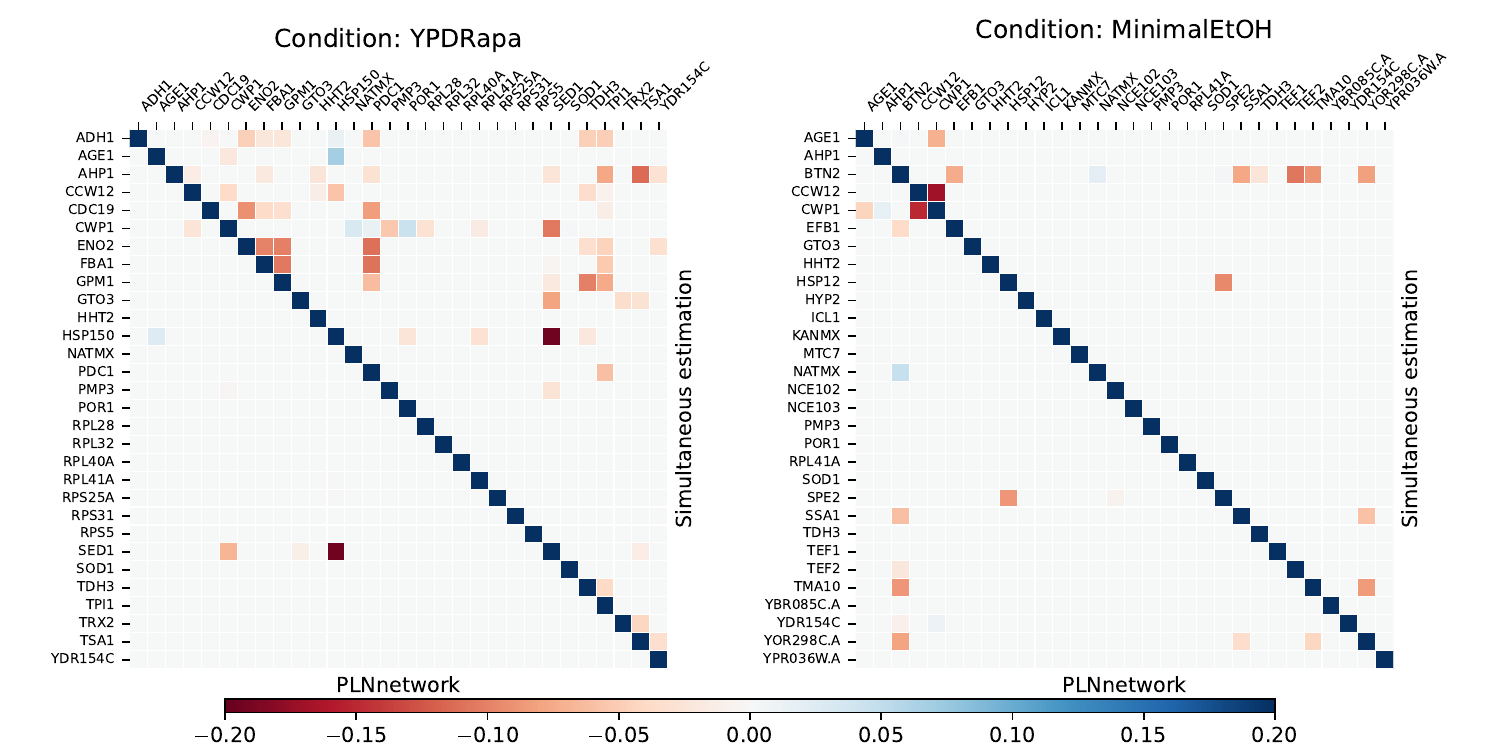}
\caption{Gene correlation matrices estimation under conditions YPDRapa and MinimalEtOH. For each condition, upper right is the result of simultaneous estimation and lower left is the result of PLNnetwork.\label{fig:matrix_main}}
    \end{subfigure}
    \caption{Yeast gene expression networks and correlation estimation results based on scRNA-seq data.}
    \end{figure}
\cite{jackson2020gene} developed a method for scRNA-seq in budding yeast using Chromium droplet-based single cell encapsulation, where they engineered transcriptional factor (TF) gene deletions by precisely excising the entire TF open reading frame and introduced a unique transcriptional barcode that enables multiplexed analysis of genotypes using scRNA-seq. They then  pooled 72 different strains, corresponding to 12 different genotypes, and determined their gene expression profiles in 11 environmental growth conditions with their respective nitrogen and carbon sources in  a total of  38,285 cells. For each cell, we have the  transcript counts for 6,834 genes. In our  analysis,  we  focus on $p=612$ genes with  expression variance greater than  1.  

We are interested in estimating the gene network of these 612 genes in each of the 11 growth conditions. We apply both the simultaneous estimation and PLNnetwork and compare the resulting estimates. For all conditions, our method gives a smaller EBIC value (Supplementary Table \plainref{tab:EBIC}), indicating a better model fitting.
Figure \ref{fig:graph_main} shows the gene expression networks for the top 70 highly expressed genes under AmmoniumSulfate condition, with networks for other conditions detailed in the Supplementary Figure \plainref{fig:graph_app}. In these visualizations, genes without any links identified by either method are omitted. The simultaneous estimation results in denser networks for highly expressed genes. This shows the proposed simultaneous estimation method results  detecting more connections for the highly expressed genes. 
Our approach also identifies hub genes that are ignored by PLNnetwork such as CWP1 and RPS22A.  This is particularly important in biological studies, as hub genes are often associated with essential biological functions and are potential targets for drug development \citep{barabasi2004network}. 

Figure \ref{fig:matrix_main} presents the scaled precision matrices for the 30 most highly expressed genes under the YPDRapa and MinimalEtOH conditions. Under YPDRapa, the precision matrix estimated by PLNnetwork is notably sparse, containing few non-zero elements. In contrast, simultaneous estimation results in a significantly denser matrix for this condition. However, under the MinimalEtOH condition, the simultaneous estimation did not produce substantial differences. This suggests that our method excels at integrating information across multiple groups when network connections are weak, while still captures task-specific patterns. Additional results for other conditions are provided in Supplementary Figure \plainref{fig:matrix_app}.

\subsection{Gene Network Analysis of T-cells in Patients with  Ulcerative Colitis}

The second dataset we analyze is  a single cell RNA-seq dataset of colon biopsies from patients with  ulcerative colitis (UC)  reported in \cite{smillie2019ibd}. UC  is characterized by chronic gastrointestinal inflammation, and  its pathogenesis is rooted in a complex interplay of genetic, environmental, and immunological factors, with T cell activity playing a pivotal role in the disease's presence \citep{powrie1995tcellibd}.  \cite{smillie2019ibd} collected the gene expression  count data of 20,529 genes from 366,650 T-cells in 18 UC patients and 12 healthy controls. For  each patient, two sets of cells   were collected, one from the inflamed colon section and another from non-inflamed colon tissue. Our analysis focuses on 13 patients who have sufficient number of T-cells in both inflamed and non-inflamed colon sections.  We further select 368 genes that  are expressed in over 30\% of these cells. We aim to understand the gene network of these 368 genes in
inflamed and non-inflamed tissues for each UC patient. 

\begin{figure}[!t]
    \centering
    \includegraphics[width=.9\linewidth]{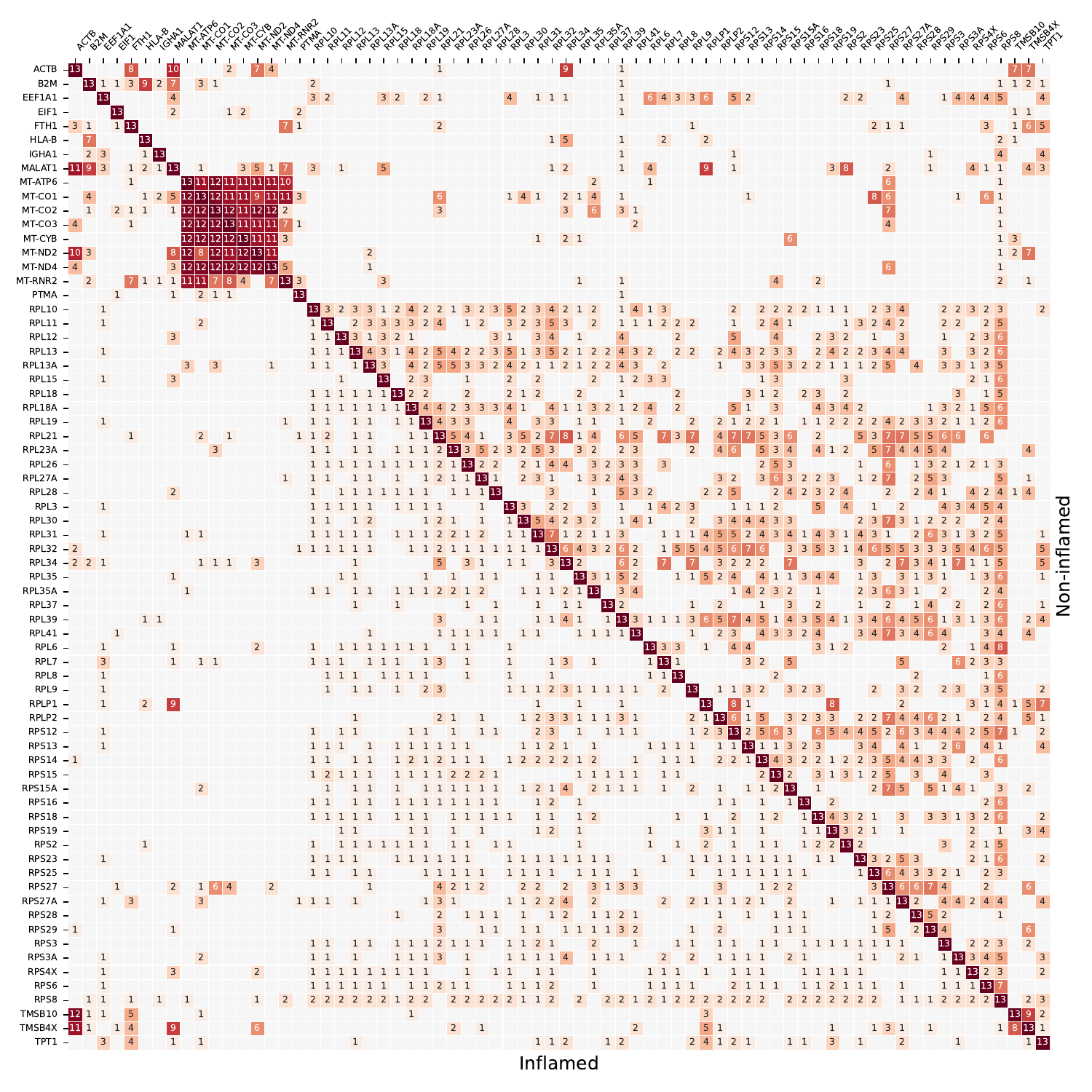}
    \caption{Comparison of gene expression networks in inflamed (lower left) vs. non-inflamed (upper right) cells of UC patients by simultaneous estimation. Each count indicates the number of patients, out of 13, having a link between each gene pair.}
    \label{fig:ibd}
\end{figure}

We apply both the simultaneous estimation method and the PLNnetwork to estimate the gene network for each of these 13 inflamed tissues, and 13 non-inflamed tissues separately and then compare the resulting networks. 
Figure \ref{fig:ibd} shows a summary of the simultaneously estimated gene expression networks for the 70 most expressed genes in both inflamed and non-inflamed T cells, marking each connection by its occurrence frequency across 13 patients. We observe two gene clusters in T cells of inflamed and non-inflamed colon tissues. The first cluster includes mitochondrially encoded (MT) genes, such as MT-CO1, MT-CYB, and MT-ND4, involved in oxidative phosphorylation. The MT gene networks in both inflamed and non-inflamed cells exhibit similar, densely connected patterns, consistently observed across the majority of the 13 patients. Mitochondria serves numerous critical cellular functions, rapidly responding to extracellular stimuli and cellular demands while dynamically communicating with other organelles. Mitochondrial function in the gastrointestinal epithelium plays a critical role in maintaining intestinal health \citep{Mit-Gene}.

The second cluster consists of ribosomal protein genes, including large ribosomal subunit (RPL) and small ribosomal subunit (RPS) genes. These genes are essential for producing a large family of proteins critical to all living cells. Ribosomal inactivation can lead to peptidyl transferase dysfunction and subsequent global translational arrest \citep{Rib-Gene}. With simultaneous estimation, we observe significant differences in the connectivity of RPL and RPS gene networks between inflamed and non-inflamed T-cells. The network in non-inflamed cells exhibits dense connectivity, with many links appearing in more than half of the patients. In contrast, the network in inflamed cells is sparsely connected, with most links observed in only 1 or 2 patients. This suggests that ribosomal protein gene activity may be suppressed in inflamed cells, consistent with previous research linking ribosomal inactivation to gut barrier disorders \citep{park2020caveolar}.

Similar to the our simultaneous estimation method, PLNnetwork also identifies clusters of MT genes and ribosomal protein genes, as shown in Supplementary Figure \plainref{fig:ibd_plnnetwork}. However, unlike our approach, PLNnetwork does not reveal significant differences between the gene networks of inflamed and non-inflamed cells, suggesting that it offers limited insight into the disease mechanism. Specifically, the connections among ribosomal protein genes in both cell types are weak, resulting in sparse networks. This further highlights the superiority of simultaneous estimation in aggregating information across samples, enhancing its effectiveness in modeling complex single-cell data.

\section{Discussion}\label{sec:discussion}
Single-cell RNA-seq data allow us to estimate condition- or sample-specific gene networks. In this paper, we have proposed  a hierarchical Poisson log-normal model for simultaneously estimating  the individual-specific  precision matrices based on the observed single cell gene expression counts. We have developed a computationally efficient variational EM algorithm to estimate the model parameters and to obtain the posterior mode of the precision matrices. We have shown in simulations that the methods lead to an overall better estimation of the network structures and the precision matrices when compared to methods that only use data from one sample. Our proposed models and methods have the flavor of the empirical Bayes methods for combining likelihoods to improve estimate of each precision matrix.  By effectively using a common prior of the precision matrices, we can effectively utilize the data observed on many samples to estimate the parameters in the prior distribution, leading to improved estimates of the precision matrices of all samples. 


\bibliographystyle{chicago}
\bibliography{ref.bib}

\newpage
\appendix

\spacingset{1.75}
\renewcommand \thefigure {\Alph{section}.\arabic{figure}}
\renewcommand \thetable {\Alph{section}.\arabic{table}}
\renewcommand \thealgorithm {\Alph{section}.\arabic{algorithm}}
\renewcommand \theequation {\Alph{section}.\arabic{equation}}
\setcounter{page}{1}
\section{Algorithms and Derivation Details}\label{supp:sec:algo_and_details}
\setcounter{table}{0}
\setcounter{figure}{0}
\setcounter{equation}{0}
\setcounter{algorithm}{0}
\subsection{Algorithms}\label{supp:sec:algo}
This section provides the ADMM update for $\bm{\mu}_{i}^{(k,t)}$ in Algorithm \ref{alg:ADMM}, and the full variational EM Algorithm \ref{alg:Omega}. Notably, all \textbf{for} loops in both algorithms are parallelizable.
\begin{algorithm}\small
\caption{ADMM algorithm for $\bm{\mu}_{i}^{(k,t)}$}
\begin{algorithmic}[1]
\renewcommand{\algorithmicrequire}{ \textbf{Initialize:}}
\renewcommand{\algorithmicensure}{ \textbf{Input:}}
\ENSURE  Count data $\bm{y}^{(k)}_i$, mean vector estimate $\bm{\lambda}^{(k,t)}_i$, variational variance estimate $\bm{\sigma}^{2(k,t-1)}_i$, precision matrix estimate $\bm{\Omega}^{(k,t)}$, convergence threshold $\epsilon$, step size $\rho$, maximum iteration number $\widetilde{T}$.
\REQUIRE $\widetilde{\Delta} > \epsilon, {\bm{\alpha}}^{(0)} = \bm{0}$, $\tilde{t}=0$,  ${\bm{\mu}}_N^{(0)}$.

\WHILE{$\tilde{t}< \widetilde{T}, \widetilde{\Delta}>\epsilon$}
\FOR{$j=1 \text{ to } p$}
\STATE $${\mu}_{M,j}^{(\tilde{t}+1)} = \argmin_{\mu_{M,j}}\left\{\ell_2(\mu_{M,j},y_{ij}^{(k)},\sigma_{ij}^{2(k,t-1)})+\alpha_i^{(\tilde{t})}(\mu_{M,j}-{\mu}_{N,j}^{(\tilde{t})})+\frac{\rho}{2}(\mu_{M,j}-{\mu}_{N,j}^{(\tilde{t})})^2\right\}.$$
\ENDFOR
\STATE Update
$${\bm{\mu}}_{N}^{(\tilde{t}+1)} = \left({\bm{\Omega}^{(k,t)}}+\rho\bm{I}\right)^{-1}\left(\rho{\bm{\mu}}_M^{(\tilde{t}+1)}+{\bm{\alpha}}^{(\tilde{t})}+\bm{\Omega}^{(k,t)}\bm{\lambda}^{(k,t)}_{i}\right).$$
\STATE Update
$${\bm{\alpha}}^{(\tilde{t}+1)} = {\bm{\alpha}}^{(\tilde{t})}+\rho\left({\bm{\mu}}_M^{(\tilde{t}+1)}-{\bm{\mu}}_N^{(\tilde{t}+1)}\right).$$
\STATE Update $\widetilde{\Delta} = \Delta ({\bm{\mu}}_M^{(\tilde{t}+1)},{\bm{\mu}}_N^{(\tilde{t}+1)})$.
\STATE $\tilde{t}=\tilde{t}+1$.
\ENDWHILE
\RETURN ${\bm{\mu}}_i^{(k,t)}={\bm{\mu}}_M^{(\tilde{t})}$.
\end{algorithmic}
\label{alg:ADMM}
\end{algorithm}
\begin{algorithm}\small
\caption{Variational EM Algorithm for Poisson Log-Normal Model}
\begin{algorithmic}[1]
\renewcommand{\algorithmicrequire}{ \textbf{Initialize:}}
\renewcommand{\algorithmicensure}{ \textbf{Input:}}
\ENSURE    Count data $\bm{y}$, offset $\bm{o}$, covariate $\bm{z}$, step size $\rho$, parameters $v_1$, $v_0$, $\tau$ and $p_0$, maximum iteration number $T>0$, convergence threshold $\epsilon$.
\REQUIRE $\Delta>\epsilon,t=0,{\bm{\Omega}}^{(\cdot,0)},{\bm{\mu}}^{(\cdot,0)},{\bm{\sigma}}^{2(\cdot,0)}$
\WHILE{$t< T, \Delta>\epsilon$}
\FOR{$i\neq j$}
\STATE Update$${p}_{ij}^{(t)} = 1/\left(1+\frac{1-p_0}{p_0}\cdot(\frac{v_1}{v_0})^K\cdot\exp\left(-(\frac{1}{v_0}-\frac{1}{v_1})\sum_{k=1}^K|{\Omega}_{ij}^{(k,t)}|)\right)\right).$$
\STATE Update$${P}_{ij}^{(t)} = \frac{{p}_{ij}^{(t)}}{v_1}+\frac{(1-{p}_{ij}^{(t)})}{v_0}.$$
\ENDFOR
\FOR{$k=1 \text{ to } K$}
\FOR{$i=1 \text{ to } n_k$}
\STATE Update ${\bm{\mu}}_{i}^{(k,t)}$ as in Algorithm \ref{alg:ADMM}.
\STATE For each $j=1,2,\cdots,p$, solve
$${\sigma}^{2(k,t)}_{ij}=\arg_{\sigma^2}\left\{\sigma^2{{\Omega}}^{(k,t)}_{jj}-\sigma^2\exp({\mu}^{(k,t)}_{ij}+\frac{1}{2}\sigma^2)=1\right\}.$$
\ENDFOR
\STATE Update ${\bm{\beta}}^{(k,t+1)} = \left({\bm{\mu}}^{(k,t)}-\bm{o}^{(k)}\right)\bm{z}^{(k)\intercal}\left(\bm{z}^{(k)}\bm{z}^{(k)\intercal}\right)^{-1}.$
\STATE $\bm{S}^{(k,t)}:=\sum_{i=1}^{n_k}\left(({\bm{\mu}}_{i}^{(k,t)}-\bm{\lambda}^{(k,t+1)}_i)({\bm{\mu}}_{i}^{(k,t)}-\bm{\lambda}^{(k,t+1)}_i)^\intercal+\mathrm{diag}({\bm{\sigma}}_{i}^{(k,t)}\right)$.
\STATE $${\bm{\Omega}}^{(k,t+1)}:=\argmax_{\bm{\Omega}}\left\{\frac{n_k}{2}\log|\bm{\Omega}|-\mathrm{Tr}\left(\bm{S}^{(k,t)}\bm{\Omega}\right)-\sum_{i=1}^p\frac{2\Omega_{ii}}{\tau}-\sum_{i< j}{P}_{ij}^{(t)}|\Omega_{ij}|
\right\}.$$
\ENDFOR
\STATE Update $\Delta = \Delta({\Omega}^{(\cdot,t)},{\Omega}^{(\cdot,t+1)})$.
\STATE $t = t+1$.
\ENDWHILE
\RETURN $\{\bm{\Omega}^{(k,t)}\}_{k=1}^K,\{\bm{\beta}^{(k,t)}\}_{k=1}^K$.
\end{algorithmic}
\label{alg:Omega}
\end{algorithm}
\subsection{Derivation of Variational EM method}\label{App.VEM}
Here we provide a detailed derivation of the variational EM method. To begin with, we introduce a set of hidden variables $\{\bm{X}^{(k)}\}_{k=1}^K$, and consider the Expectation step. We express the full hierarchical PLN model as follows:
\begin{equation}
    \begin{gathered}
        \bm{\Gamma}\sim\prod_{i<j}p_0^{\gamma_{ij}}(1-p_0)^{1-\gamma_{ij}},\\
        \bm{\Omega}^{(k)}|\bm{\Gamma}\overset{ind}{\sim} \I_{\{\bm{\Omega}^{(k)}>0\}}\cdot\prod_{i<j}\left(\frac{1}{2v_{\gamma_{ij}}}\exp(-|\Omega_{ij}^{(k)}|/v_{\gamma_{ij}})\right)\cdot\prod_{i=1}^p\frac{1}{\tau}\exp(-\Omega_{ii}^{(k)}/\tau),\\
        \bm{X}_i^{(k)}|\bm{\Omega}^{(k)},\bm{\beta}^{(k)}\overset{ind}{\sim} N(\bm{\lambda}_i^{(k)},(\bm{\Omega}^{(k)})^{-1}),\\
        y_{ij}^{(k)}|X_{ij}^{(k)}\sim\mathrm{Pois}(\exp(X_{ij}^{(k)})).
    \end{gathered}
\end{equation}
Given the current estimation $\{\bm{\Omega}^{(k,t)}\}_{k=1}^K$ and $\{\bm{\beta}^{(k,t)}\}_{k=1}^K$, the expected value of the log-likelihood function is as follows:
\begin{equation}
    \begin{aligned}
        {\ell}(\bm{\Omega},\bm{\beta}|\bm{\Omega}^{(\cdot,t)},\bm{\beta}^{(\cdot,t)})&=\E_{\bm{X}\sim\Prob(\cdot|\bm{\Omega}^{(\cdot,t)},\bm{\beta}^{(\cdot,t)})}\sum_{k=1}^K\log\Prob(\bm{y}^{(k)},\bm{X}^{(k)}|\bm{\Omega}^{(k)},\bm{\beta}^{(k)})\\
        &=\E_{\bm{X}\sim\Prob(\cdot|\bm{\Omega}^{(\cdot,t)},\bm{\beta}^{(\cdot,t)})}\sum_{k=1}^K\log\Prob(\bm{X}^{(k)}|\bm{\Omega}^{(k)},\bm{\beta}^{(k)})\Prob(\bm{y}^{(k)}|\bm{X}^{(k)})\\
        &=\E_{\bm{X}\sim\Prob(\cdot|\bm{\Omega}^{(\cdot,t)},\bm{\beta}^{(\cdot,t)})}\sum_{k=1}^K\log\Prob(\bm{X}^{(k)}|\bm{\Omega}^{(k)},\bm{\beta}^{(k)})+C\\
        &=\E_{\bm{X}\sim\Prob(\cdot|\bm{\Omega}^{(\cdot,t)},\bm{\beta}^{(\cdot,t)})}\sum_{k=1}^K\sum_{i=1}^{n_k}\left(\frac{1}{2}\log|\bm{\Omega}^{(k)}|-\frac{1}{2}(\bm{X}_i^{(k)}-\bm{\lambda}_{i}^{(k)})^\intercal\bm{\Omega}^{(k)}(\bm{X}_i^{(k)}-\bm{\lambda}_{i}^{(k)})\right)+C,
    \end{aligned}
\end{equation}
where $C$ is a constant unrelated to $(\bm{\Omega}, \bm{\beta})$ and may vary between different lines. For the above equation, the second line is derived by the Bayes rule, and the last line is due to the independence and normality of $\bm{X}_i^{(k)}$ given $\bm{\Omega}^{(k)}$ and $\bm{\beta}^{(k)}$. By replacing the posterior of $\bm{X}$ with the variational posterior $\widetilde{\Pi}(\cdot,t)$ , we can write the variational surrogate of the log-likelihood as:
\begin{equation}
    \begin{aligned}
        \widetilde{\ell}(\bm{\Omega},\bm{\beta}|\bm{\Omega}^{(\cdot,t)},\bm{\beta}^{(\cdot,t)})&=\sum_{k=1}^K\sum_{i=1}^{n_k}\left(\frac{1}{2}\log|\bm{\Omega}^{(k)}|-\frac{1}{2}({\bm{\mu}}_{i}^{(k,t)}-\bm{\lambda}^{(k)}_i)^\intercal\bm{\Omega}^{(k)}({\bm{\mu}}_{i}^{(k,t)}-\bm{\lambda}^{(k)}_i)\right.\\
        &-\left.\frac{1}{2}\mathrm{Tr}(\mathrm{diag}(\bm{\sigma}_i^{2(k,t)})\bm{\Omega}^{(k)})\right)+C,
    \end{aligned}
    \label{eq:vi-loglik}
\end{equation}
where $\bm{\mu}_{i}^{(k,t)}$ and $\bm{\sigma}_i^{2(k,t)}$ are the parameters in the variational family and we'll show how to get their expression in the end of this section. Next, we consider the prior part of the EM algorithm. By letting $\bm{\Gamma}$ be another hidden variable, the expected log prior is given by:
\begin{equation}
    \begin{aligned}
        \E_{\bm{\Gamma}\sim\Prob(\cdot|\bm{\Omega}^{(\cdot,t)})}\log\Prob({\bm{\Omega},\bm{\Gamma}})&=\E_{\bm{\Gamma}\sim\Prob(\cdot|\bm{\Omega}^{(\cdot,t)})}\log\prod_{k=1}^K\Prob({\bm{\Omega}^{(k)}|\bm{\Gamma}})\cdot\Prob(\bm{\Gamma})\\
        &=\E_{\bm{\Gamma}\sim\Prob(\cdot|\bm{\Omega}^{(\cdot,t)})}\log\prod_{k=1}^K\Prob({\bm{\Omega}^{(k)}|\bm{\Gamma}})+C\\
        &=\E_{\bm{\Gamma}\sim\Prob(\cdot|\bm{\Omega}^{(\cdot,t)})}\sum_{k=1}^K\left(-\sum_{i<j}|\Omega_{ij}^{(k)}|/2{v_{\gamma_{ij}}}-\sum_{i=1}^{p}\Omega_{ii}^{(k)}/\tau\right)+C \\
        &=-\sum_{k=1}^K\left(\sum_{i<j}(\frac{p_{ij}^{(t)}}{2v_1}+\frac{1-p_{ij}^{(t)}}{2v_0})|\Omega^{(k)}_{ij}|+\sum_{i=1}^{p}\frac{1}{\tau}\Omega_{ii}^{(k)}\right)+C.
    \end{aligned}
    \label{eq:vi-prior}
\end{equation}
For the last equation, $p_{ij}^{(t)}=\Prob(\gamma_{ij}=1|\bm{\Omega}^{(\cdot,t)})$ is the posterior probability of $\gamma_{ij}=1$ conditioning on $\bm{\Omega}^{(\cdot,t)}$, which we'll define later. Combining \eqref{eq:vi-loglik} and \eqref{eq:vi-prior}, we get the variational log posterior function:
\begin{equation}
    \widetilde{Q}(\bm{\Omega},\bm{\beta}|\bm{\Omega}^{(\cdot,t)},\bm{\beta}^{(\cdot,t)})=\frac{1}{2}\sum_{k=1}^K\left(n_k\log|\bm{\Omega}^{(k)}|-\mathrm{Tr}(\widetilde{\bm{S}}^{(k,t)}\bm{\Omega}^{(k)})-\sum_{i<j}P_{ij}^{(t)}|\Omega_{ij}^{(k)}|-\sum_{i=1}^{p}\frac{2\Omega_{ii}^{(k)}}{\tau}\right),
    \label{eq:tilde_Q}
\end{equation}
where
\begin{equation}
    \widetilde{\bm{S}}^{(k,t)}=\sum_{i=1}^{n_k}\left(({\bm{\mu}}_{i}^{(k,t)}-\bm{\lambda}^{(k)}_i)({\bm{\mu}}_{i}^{(k,t)}-\bm{\lambda}^{(k)}_i)^\intercal+\mathrm{diag}({\bm{\sigma}}_{i}^{(k,t)})\right)
\end{equation}
is the variational covariance matrix and
\begin{equation}
    {P}_{ij}^{(t)} = \frac{{p}_{ij}^{(t)}}{v_1}+\frac{(1-{p}_{ij}^{(t)})}{v_0}
\end{equation}
is the penalty term of the $(i,j)$-th element for every group. Notably, equation \eqref{eq:tilde_Q} is separable for the $K$ groups, encouraging simultaneous optimization for different groups in the Maximization step.

To get the update rule for $\bm{\beta}^{(k)}$, we setting its partial derivative in \eqref{eq:tilde_Q} to zero, which can be computed as
\begin{equation}
\begin{aligned}
        &\frac{\partial}{\partial\bm{\beta}^{(k)}}\sum_{i=1}^{n_k}\left(\bm{\mu}_i^{(k,t)}-\bm{o}^{(k)}-\bm{\beta}^{(k)}\bm{z}_{i}^{(k)}\right)^\intercal\bm{\Omega}^{(k)}\left(\bm{\mu}_i^{(k,t)}-\bm{o}^{(k)}-\bm{\beta}^{(k)}\bm{z}_{i}^{(k)}\right)\\
        =&2\bm{\Omega}^{(k)}\left(\bm{\beta}^{(k)}\bm{z}^{(k)}\bm{z}^{(k)\intercal}-\bm{\mu}^{(k,t)}\bm{z}^{(k)\intercal}-\bm{o}^{(k)}\bm{z}^{(k)\intercal}\right)=0,
\end{aligned}
\end{equation}
leading to
\begin{equation}
    {\bm{\beta}}^{(k,t+1)} = \left({\bm{\mu}}^{(k,t)}-\bm{o}^{(k)}\right)\bm{z}^{(k)\intercal}\left(\bm{z}^{(k)}\bm{z}^{(k)\intercal}\right)^{-1}
\end{equation}
and
\begin{equation}
  \bm{\Omega}^{(k,t+1)}=\argmax_{\bm{\Omega}^{(k)}}\left\{n_k\log|\bm{\Omega}^{(k)}|-\mathrm{Tr}({\bm{S}}^{(k,t)}\bm{\Omega}^{(k)})-\sum_{i<j}P_{ij}^{(t)}|\Omega_{ij}^{(k)}|-\sum_{i=1}^{p}\frac{2\Omega_{ii}^{(k)}}{\tau}\right\},
\end{equation}
where
\begin{equation}
    \begin{aligned}
            {\bm{S}}^{(k,t)}&=\sum_{i=1}^{n_k}\left(({\bm{\mu}}_{i}^{(k,t)}-\bm{\lambda}^{(k,t+1)}_i)({\bm{\mu}}_{i}^{(k,t)}-\bm{\lambda}^{(k,t+1)}_i)^\intercal+\mathrm{diag}({\bm{\sigma}}_{i}^{(k,t)})\right),\\
            \bm{\lambda}^{(k,t+1)}_i&=\bm{o}_i^{(k)}+\bm{\beta}^{(k,t+1)}\bm{z}_i^{(k)}.
    \end{aligned}
\end{equation}
Next, let's derive the posterior inclusion probability $p_{ij}^{(t)}$ and the variational posterior of $\bm{X}$. Employing Bayes' rule, we compute the posterior inclusion as:
\begin{equation}
    \begin{aligned}
        p_{ij}^{(t)}&=\Prob(\gamma_{ij}=1|\bm{\Omega}^{(\cdot,t)})=\Prob(\bm{\Omega}^{(\cdot,t)}|\gamma_{ij}=1)\Prob(\gamma_{ij}=1)/\Prob(\bm{\Omega}^{(\cdot,t)})\\
        &\propto\Prob(\bm{\Omega}^{(\cdot,t)}_{ij}|\gamma_{ij}=1)\Prob(\gamma_{ij}=1)\\
        &=\prod_{k=1}^K\Prob(\Omega^{(\cdot,t)}_{ij}|\gamma_{ij}=1)\cdot\Prob(\gamma_{ij}=1)\\
        &\propto \prod_{k=1}^K\left(\frac{1}{v_1}\exp(-|\Omega_{ij}^{(k,t)}|/v_1)\right)\cdot p_0\\
        &=\exp\left(-\|\bm{\Omega}^{(\cdot,t)}_{ij}\|_1/v_1\right)\frac{p_0}{v_1^K}.
    \end{aligned}
\end{equation}
Likewise, $\Prob(\gamma_{ij}=0|\bm{\Omega}^{(\cdot,t)}_{ij})\propto\exp\left(-\|\bm{\Omega}^{(\cdot,t)}\|_0/v_0\right)\frac{1-p_0}{v_0^K}$. It's important to note that $\Prob(\gamma_{ij}=1|\bm{\Omega}^{(\cdot,t)})+\Prob(\gamma_{ij}=0|\bm{\Omega}^{(\cdot,t)})=1$. Therefore, we have:
\begin{equation}
    p_{ij}^{(t)}=\frac{1}{1+\frac{1-p_0}{p_0}(\frac{v_1}{v_0})^K\exp\left(-(\frac{1}{v_0}-\frac{1}{v_1})\|\bm{\Omega}^{(\cdot,t)}_{ij}\|_1\right)}.
\end{equation}
This formula provides the posterior inclusion probability for $\gamma_{ij}$ given $\bm{\Omega}^{(\cdot,t)}$.

It remains to give the expression of the variational posterior. Note that in the mean-field family, the variational distributions of $X_{ij}^{(k)}$ are independent for every $i,j$ and group $k$. Furthermore, given $\bm{y}$, $\bm{\Omega}^{(\cdot,t)}$ and $\bm{\beta}^{(\cdot,t)}$, the conditional distribution of $\bm{X}$ can also be expressed as an independent product:
\begin{equation}
   \Prob(\bm{X}|\bm{y},\bm{\Omega}^{(\cdot,t)},\bm{\beta}^{(\cdot,t)})=\prod_{k=1}^K\prod_{i=1}^{n_k}\Prob(\bm{X}_{i}^{(k)}|\bm{y}_{i}^{(k)},\bm{\Omega}^{(k,t)},\bm{\beta}^{(k,t)}).
\end{equation}
Therefore, the KL divergence can be written as the following summation:
\begin{equation}
    D_{\text{KL}}(P(\bm{X})\|\Prob(\bm{X}|\bm{y},\bm{\Omega}^{(\cdot,t)},\bm{\beta}^{(\cdot,t)}))=\sum_{k=1}^K\sum_{i=1}^{n_k}D_{\text{KL}}(P(\bm{X}^{(k)}_i)\|\Prob(\bm{X}^{(k)}_i|\bm{y}^{(k)}_i,\bm{\Omega}^{(k,t)},\bm{\beta}^{(k,t)})).
\end{equation}
Next we figure out the posterior probability $\Prob(\bm{X}_{i}^{(k)}|\bm{y}_{i}^{(k)},\bm{\Omega}^{(k,t)},\bm{\beta}^{(k,t)})$, which is given by:
\begin{equation}
\begin{aligned}
    \Prob(\bm{X}_{i}^{(k)}|\bm{y}_{i}^{(k)},\bm{\Omega}^{(k,t)},\bm{\beta}^{(k,t)})&=\Prob(\bm{X}_{i}^{(k)},\bm{y}_{i}^{(k)},\bm{\Omega}^{(k,t)},\bm{\beta}^{(k,t)})/\Prob(\bm{y}_{i}^{(k)},\bm{\Omega}^{(k,t)},\bm{\beta}^{(k,t)})\\
    &=\Prob(\bm{X}_{i}^{(k)}|\bm{\Omega}^{(k,t)},\bm{\beta}^{(k,t)})\Prob(\bm{y}_{i}^{(k)}|\bm{X}_{i}^{(k)})/\Prob(\bm{y}_{i}^{(k)},\bm{\Omega}^{(k,t)},\bm{\beta}^{(k,t)})\\
    &\propto N(\bm{X}_{i}^{(k)};\bm{\lambda}_i^{(k)},(\bm{\Omega}^{(k)})^{-1})\prod_{j=1}^{p}\exp\left(X_{ij}^{(k)}y_{ij}^{(k)}-\exp(X_{ij}^{(k)})\right).
\end{aligned}
\end{equation}
Plugging in this posterior probability, the KL divergence has the following form:
\begin{equation}
    \begin{aligned}
        D_{\text{KL}}(P(\bm{X}_{i}^{(k)})\|\Prob(\bm{X}_{i}^{(k)}|\bm{y}_i^{(k)},\bm{\Omega}^{(k,t)},\bm{\beta}^{(k,t)}))
        &=D_{\text{KL}}\left(N(\bm{\mu}_{i}^{(k)},\mathrm{diag}(\bm{\sigma}^{2(k)}_{i}))||N(\bm{\lambda}_i^{(k)},(\bm{\Omega}^{(k)})^{-1})\right)\\
        &-\sum_{j=1}^p\E\left[X_{ij}^{(k)}y_{ij}^{(k)}-\exp(X_{ij}^{(k)})\right]\\
        &=\frac{1}{2}(\bm{\mu}_i^{(k)}-\bm{\lambda}^{(k)}_{i})^\intercal\bm{\Omega}^{(k,t)}(\bm{\mu}_i^{(k)}-\bm{\lambda}^{(k)}_{i})-\bm{y}^{(k)\intercal}_i\bm{\mu}_i^{(k)}\\
    &+\frac{1}{2}\sum_{j=1}^p\left(\sigma_{ij}^{2(k)}\Omega_{jj}^{(k,t)}+2\exp(\mu_{ij}^{(k)}+\frac{\sigma^{2(k)}_{ij}}{2})-\log\sigma_{ij}^{2(k)}\right).
    \end{aligned}
\end{equation}
The minimization of the KL divergence with respect to $\bm{\sigma}^{2(k)}_{i}$ is separable for each coordinate, allowing the use of univariate optimization algorithms to perform simultaneous optimization. The update for $\bm{\mu}_i^{(k)}$ is obtained using the ADMM algorithm, which has a Lagrangian
\begin{equation}
    \begin{aligned}
        &\frac{1}{2}(\bm{\mu}_N-\bm{\lambda}_i^{(k)})^\intercal\bm{\Omega}^{(k,t)}(\bm{\mu}_N-\bm{\lambda}_i^{(k)})-\bm{y}_i^{(k)\intercal}\bm{\mu}_M\\
        +&\frac{1}{2}\sum_{j=1}^p\left(\sigma_{ij}^{2(k)}\Omega_{jj}^{(k,t)}+2\exp(\mu_{ij}^{(k)}+\frac{\sigma^{2(k)}_{ij}}{2})-\log\sigma_{ij}^{2(k)}\right)
    \end{aligned}
\end{equation}

\section{Numerical Study and Additional Results}\label{supp:sec:additional_results}
\setcounter{table}{0}
\setcounter{figure}{0}
\setcounter{equation}{0}
\setcounter{algorithm}{0}
\subsection{Simulation Setup}\label{App.simu}
For our method, we initialize the variational mean ${\bm{\mu}}^{(k,0)}=\log(\bm{y}^{(k)}+0.5)$ and variational variance ${\bm{\sigma}^2}^{(k,0)}=1.1\times\bm{1}$ for $k=1,2,\cdots,K$. The initial value of precision matrix ${\bm{\Omega}}^{(k,0)}$ is set as $\left({\bm{\mu}}^{(k,0)}{\bm{\mu}}^{(k,0)\intercal}/n_k+0.01\bm{I}_p\right)^{-1}$.
For Section \plainref{sec:simulation_pln}, the precision matrices are generated according to the following scheme: $\bm{\Omega}^{(k)}=(\bm{A}^{(k)} + \bm{I}_{p})\odot \bm{B}^{(k)} +\epsilon^{(k)}\bm{I}_p$, where $\bm{A}^{(k)}\in\{0,1\}^{p\times p}$ is the adjacency matrix of the $k$th group, $\bm{B}^{(k)}=(b_{ij}^{(k)})_{p\times p}$ is a symmetric random matrix with independent elements $b^{(k)}_{ii} \sim \mathrm{Unif}(0.8,1.2)$ and $b^{(k)}_{ij}\sim \mathrm{Unif}([-0.5,-0.2]\cup[0.2,0.5]),i<j$, and the symbol $\odot$ denotes the element-wise (Hadamard) product of two matrices, i.e. $(x_{ij})_{p\times p}\odot (y_{ij})_{p\times p} = (x_{ij}y_{ij})_{p\times p}$. The parameter $\epsilon^{(k)}$ is determined as $\max\left\{-\lambda_{\min}(\bm{A}^{(k)} + \bm{I}_{p})\odot \bm{B}^{(k)},0\right\}+0.01$ to ensure that $\bm{\Omega}^{(k)}$ remains positive definite. For each group $k$, we set the offset matrix $\bm{o}^{(k)}$ to zero, and the coefficients are generated independently by $\bm{\beta}_j^{(k)}\sim N(\bm{0}_p,\bm{I}_p)$. 

In Section \plainref{sec:simulation_pln_various}, the precision matrices have the similar structures, except that the distribution of nonzero elements are different: the matrix $\bm{B}^{(k)}$ have independent elements $b_{ii}^{(k)}\sim \mathrm{Unif(1,1.5)}$ and $b_{ij}^{(k)}\sim \mathrm{Unif}([-0.8,-0.3]\cup[0.3,0.8])\times s_g, i<j$. The generation of both coefficients and samples follows the same process as in Section \plainref{sec:simulation_pln}.
\subsection{Parallel Computing}\label{supp:sec:parallel}
\begin{figure}[!ht]
\centering
\includegraphics[width=0.6\linewidth]{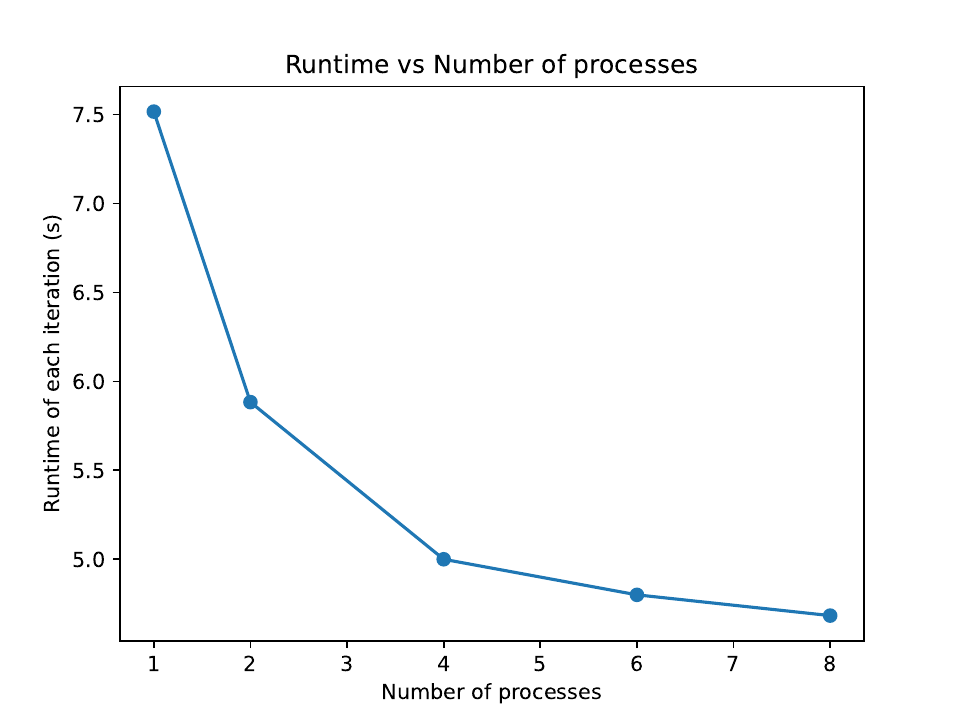}
    \caption{Runtime of simultaneous estimation using different numbers of processes.}
    \label{fig:computation}
    \end{figure}

Here we evaluate the computational efficiency of simultaneous estimation with parallel processing, where we analyze the running time across different processes. The simulation setting is the same as Section \plainref{sec:simulation_pln}, with $n=500, p=40, K=30, s=0.8$. The experiment is conducted in a MacBook Pro with M1 Pro and 16GB memory. The result is given in Figure \ref{fig:computation}, where we observe parallel processing significantly reduces computation time, especially in comparison to using a single process.

\ignore{
\begin{figure}[ht]
    \centering
    \includegraphics[width=0.9\columnwidth]{figure/app_lasso.pdf}
    \caption{The result for Variable Selection under model misspecification.}
    \label{fig:multinomial}
\end{figure}}

\subsection{Supporting Figures and Tables}\label{supp:sec:supporting}
This section contains supporting figures and tables for the main text, including simulation and real data analysis.

\begin{table}[h]
\centering
\begin{tabular}{l|r|r}
\toprule
                & Our method (Simultaneous) & PLNnetwork \\ \hline
Glutamine       & 8.58       & 8.69       \\
Proline         & 3.12       & 3.26       \\
AmmoniumSulfate & 5.05       & 5.12       \\
Urea            & 4.45       & 4.54       \\
YPD             & 58.51      & 58.63      \\
YPDRapa         & 55.52      & 55.72      \\
CStarve         & 8.25       & 8.38       \\
MinimalGlucose  & 10.37      & 10.50      \\
MinimalEtOH     & 2.33       & 2.41       \\
YPDDiauxic      & 8.34       & 8.49       \\
YPEtOH          & 6.99       & 7.05       \\ \bottomrule
\end{tabular}
\caption{EBIC($\times 10^{-5}$) of our method (Simultaneous) and PLNnetwork in eleven conditions for the yeast Single cell Data analysis.}
\label{tab:EBIC}
\end{table}

\begin{figure}[!ht]
    \centering
    \includegraphics[width=0.9\columnwidth]{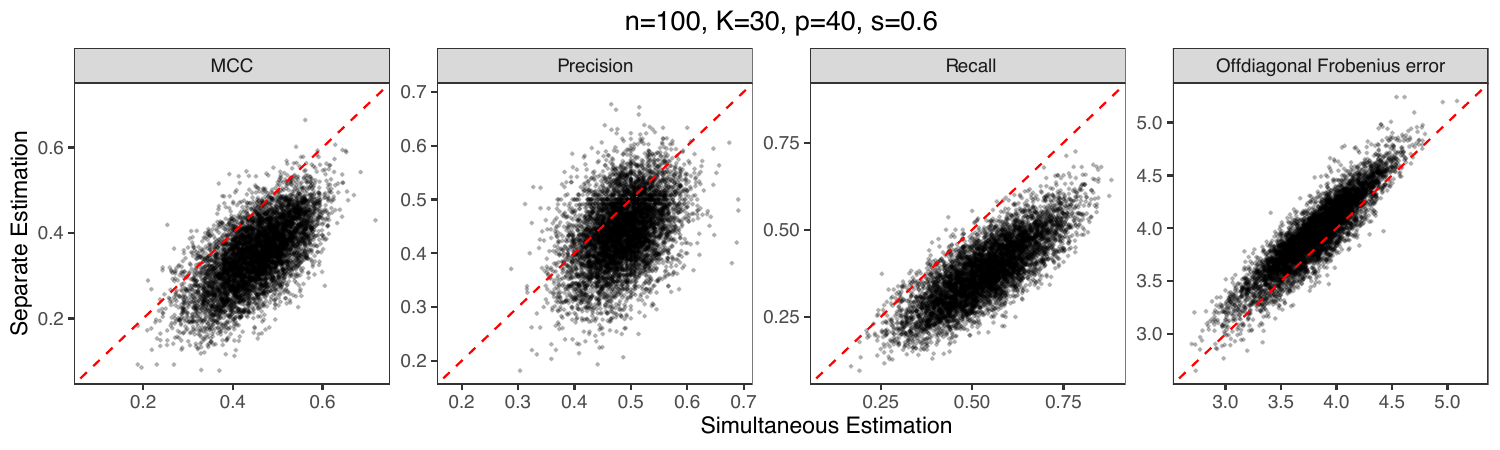}
    \caption{Individual level comparison of simultaneous estimation and separate estimation for Erdős–Rényi graphs. Red dashed lines indicate equal evaluations. Each dot represents the result of a network.}
    \label{fig:comp_individual}
\end{figure}


\begin{figure}
\centering
\begin{subfigure}[b]{.8\linewidth}
\includegraphics[width=\linewidth]{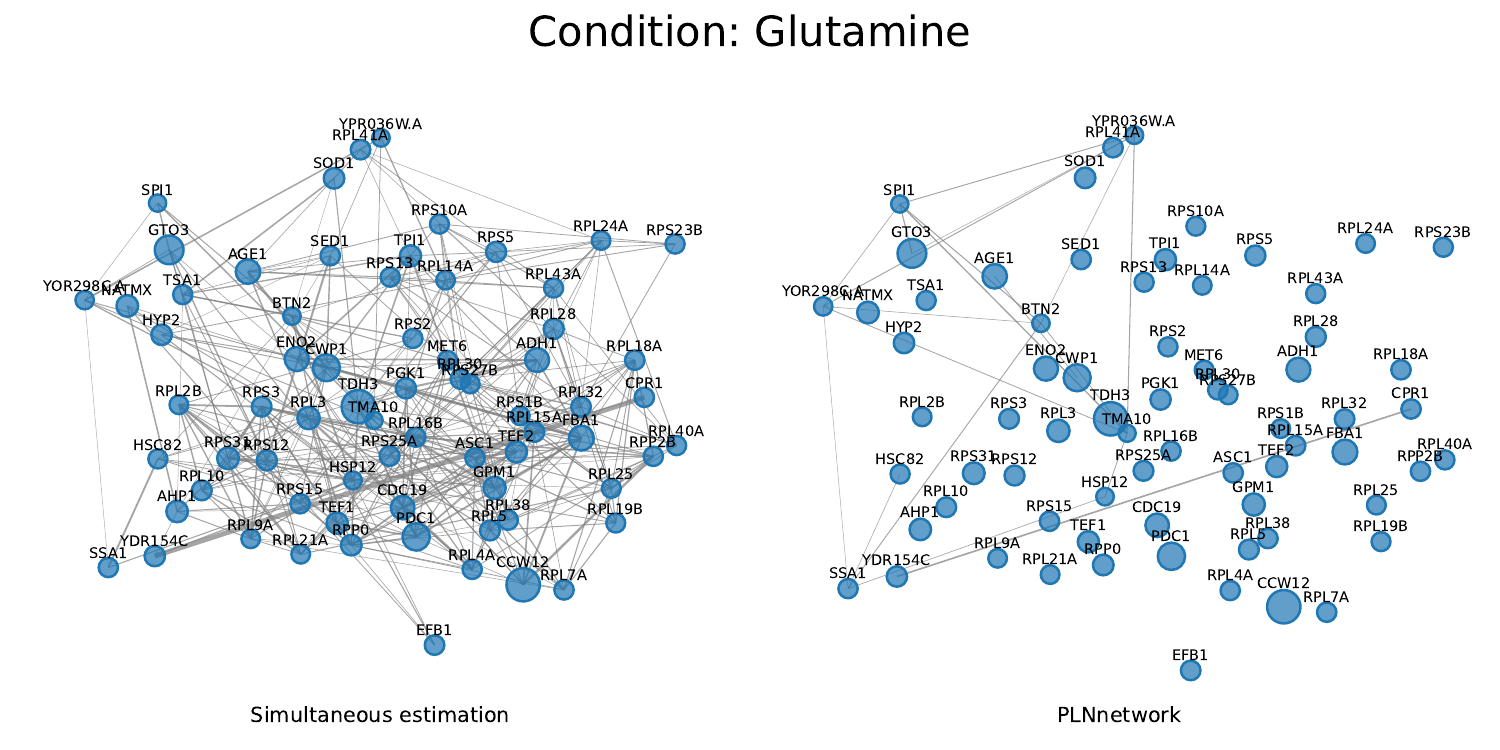}
\end{subfigure}
\begin{subfigure}[b]{.8\linewidth}
\includegraphics[width=\linewidth]{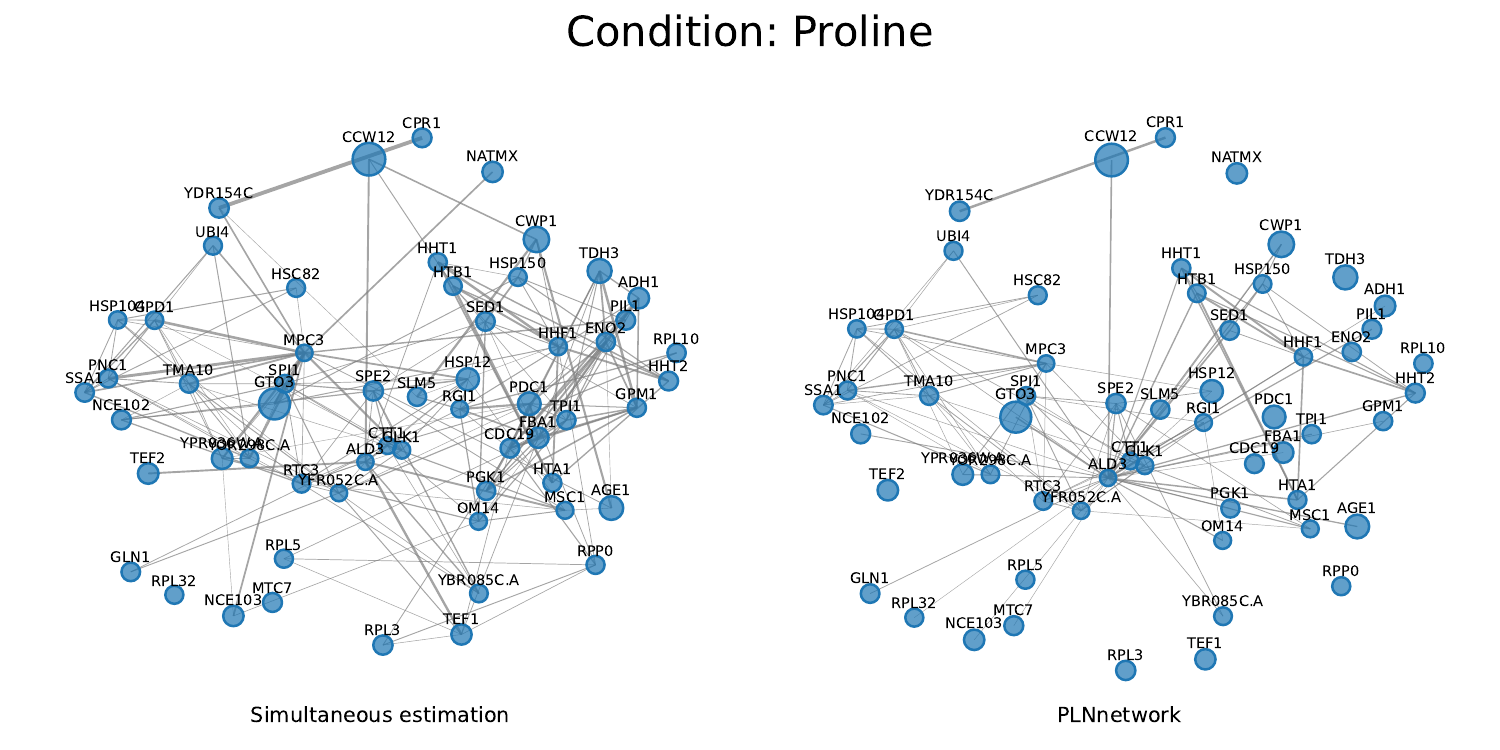}
\end{subfigure}
\begin{subfigure}[b]{.8\linewidth}
\includegraphics[width=\linewidth]{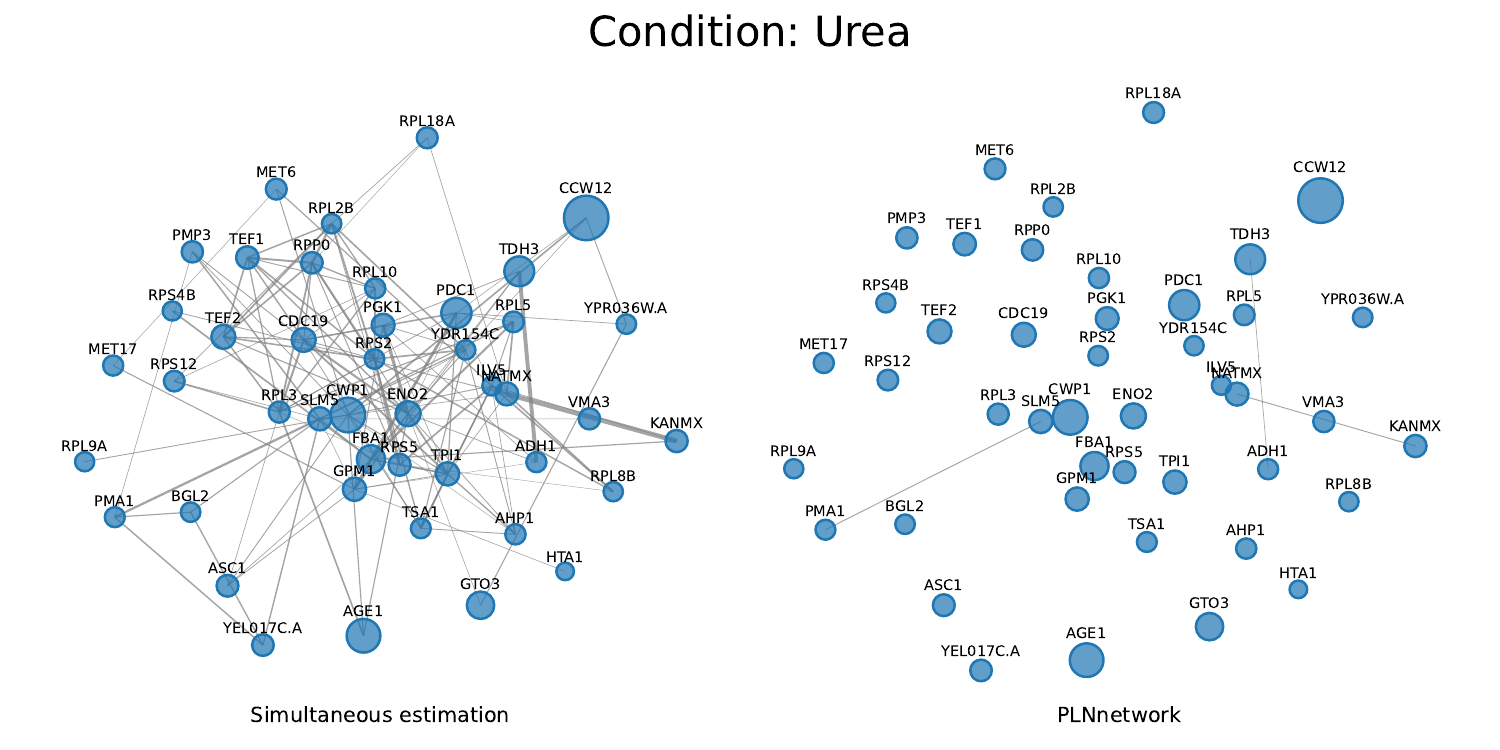}
\end{subfigure}
\caption{Gene expression network estimation for various conditions. For each condition, left is estimated by our method and right is by PLNnetwork. The legend is the same as Figure \plainref{fig:graph_main}.}
\label{fig:graph_app}
\end{figure}

\clearpage
\begin{figure}
    \centering
    \ContinuedFloat
\begin{subfigure}[b]{.8\linewidth}
\includegraphics[width=\linewidth]{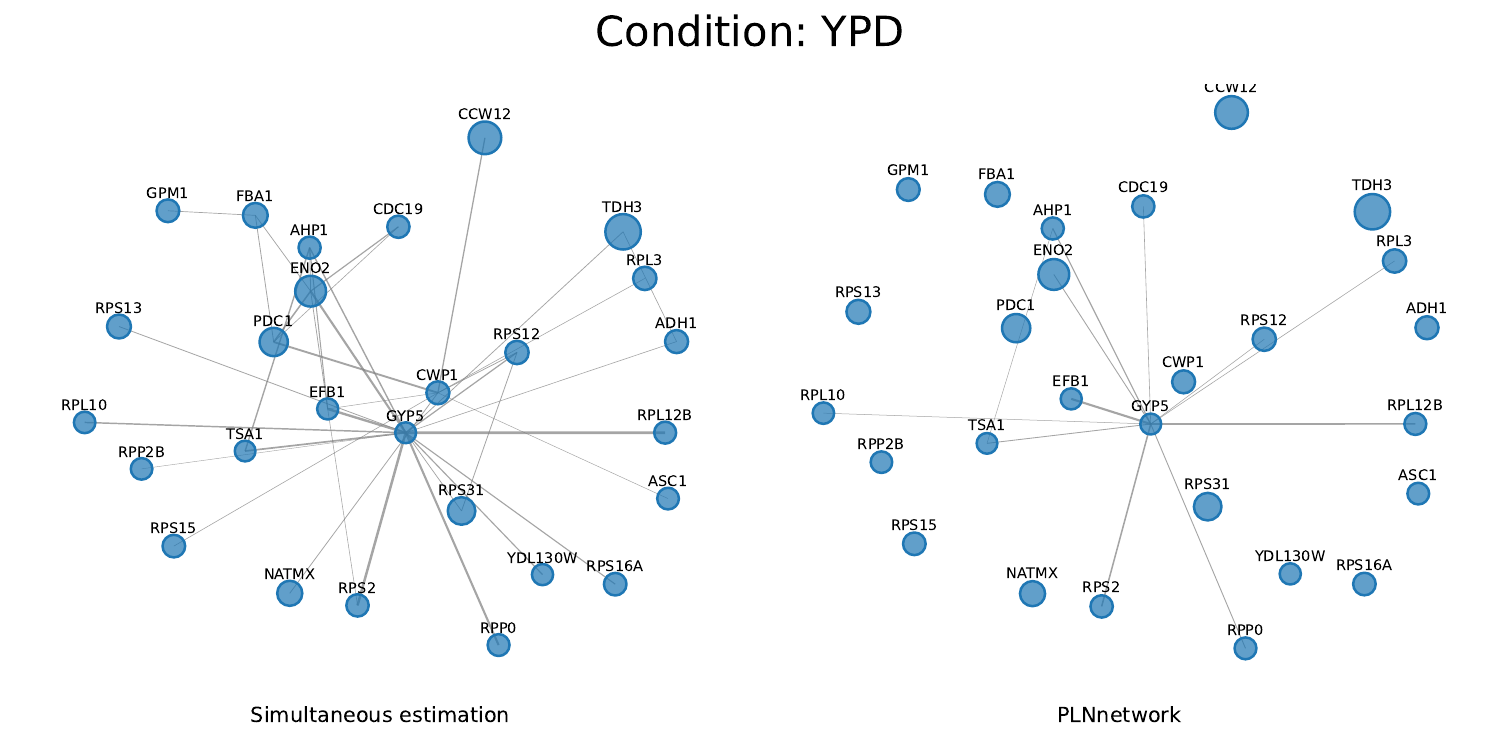}
\end{subfigure}
\begin{subfigure}[b]{.8\linewidth}
\includegraphics[width=\linewidth]{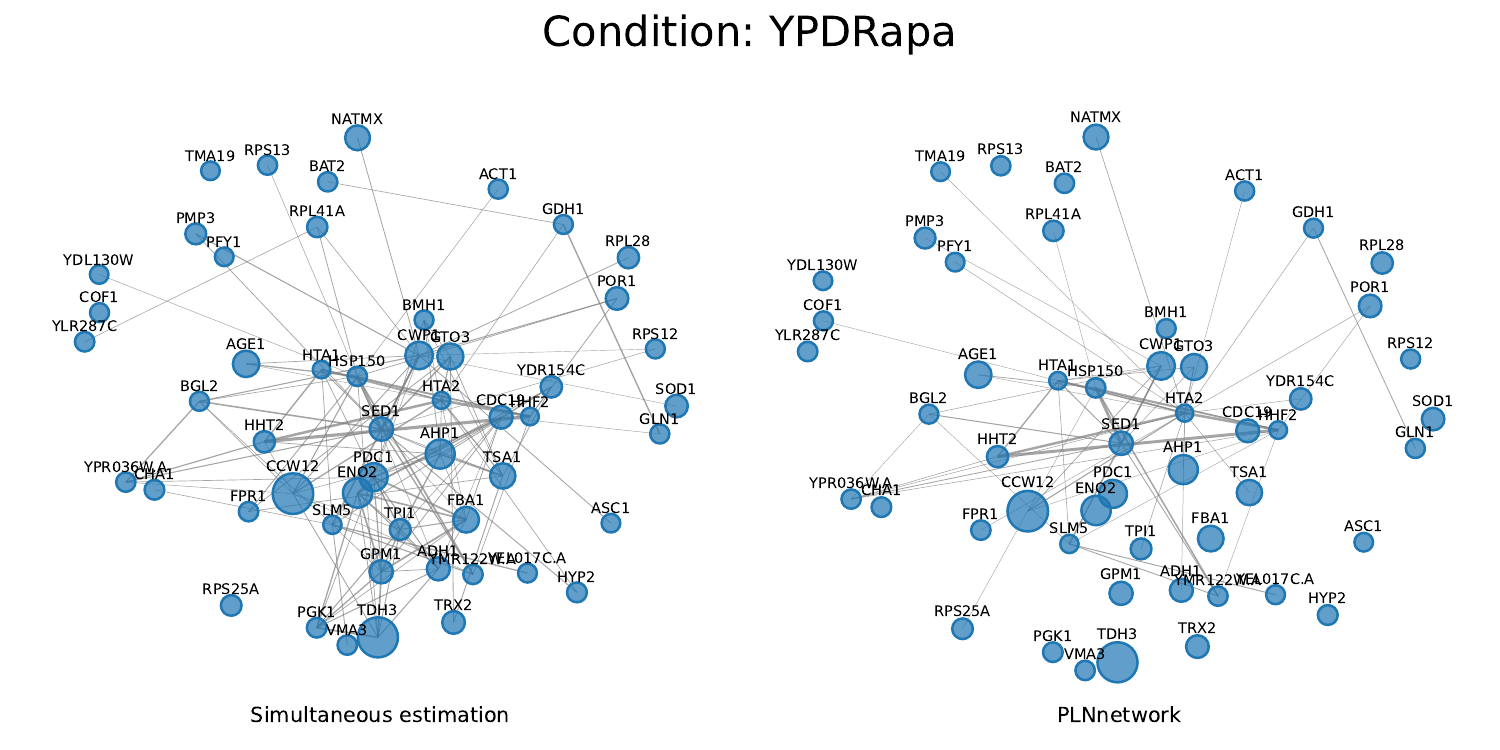}
\end{subfigure}
\begin{subfigure}[b]{.8\linewidth}
\includegraphics[width=\linewidth]{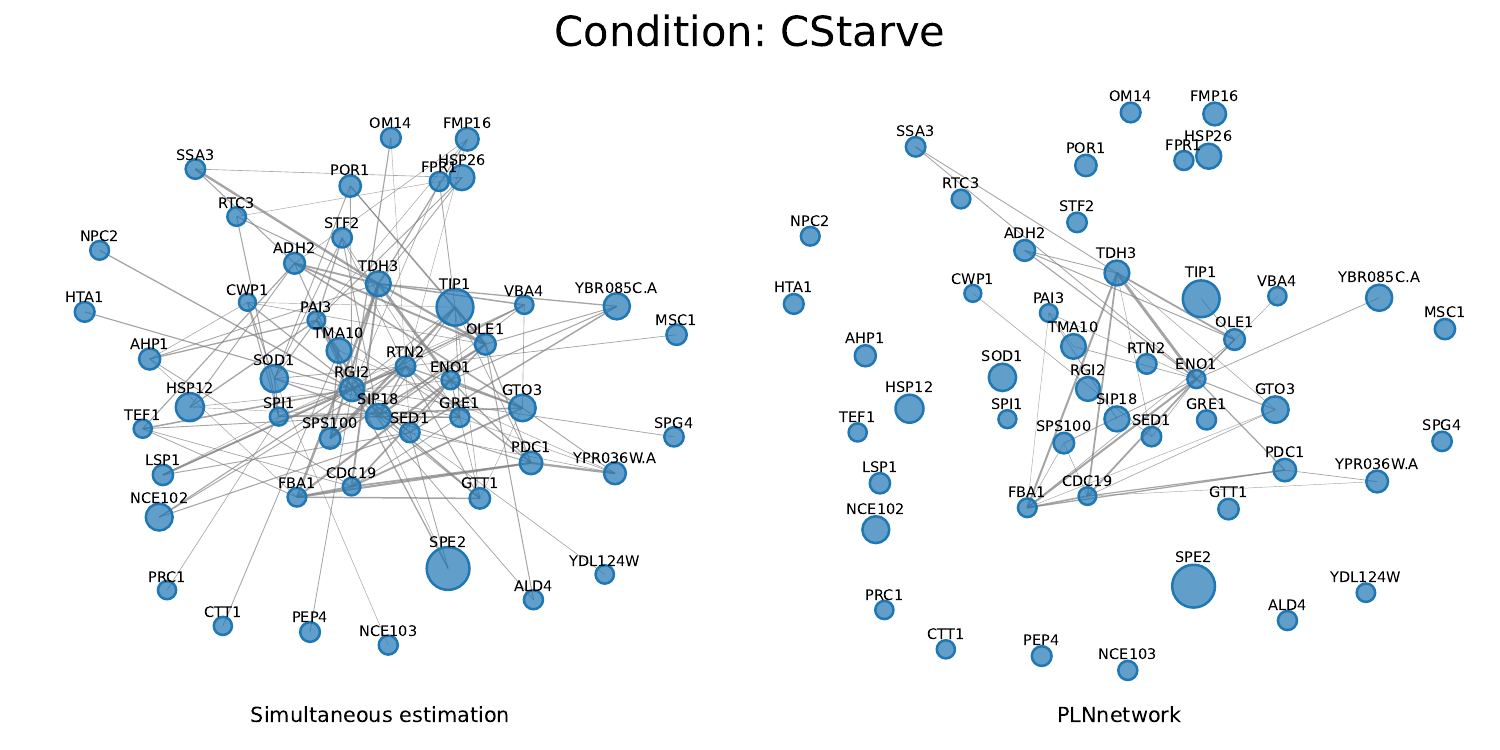}
\end{subfigure}
\caption{Gene expression network estimation for various conditions (cont.)}
\end{figure}

\clearpage
\begin{figure}
\centering
\ContinuedFloat
\begin{subfigure}[b]{.8\linewidth}
\includegraphics[width=\linewidth]{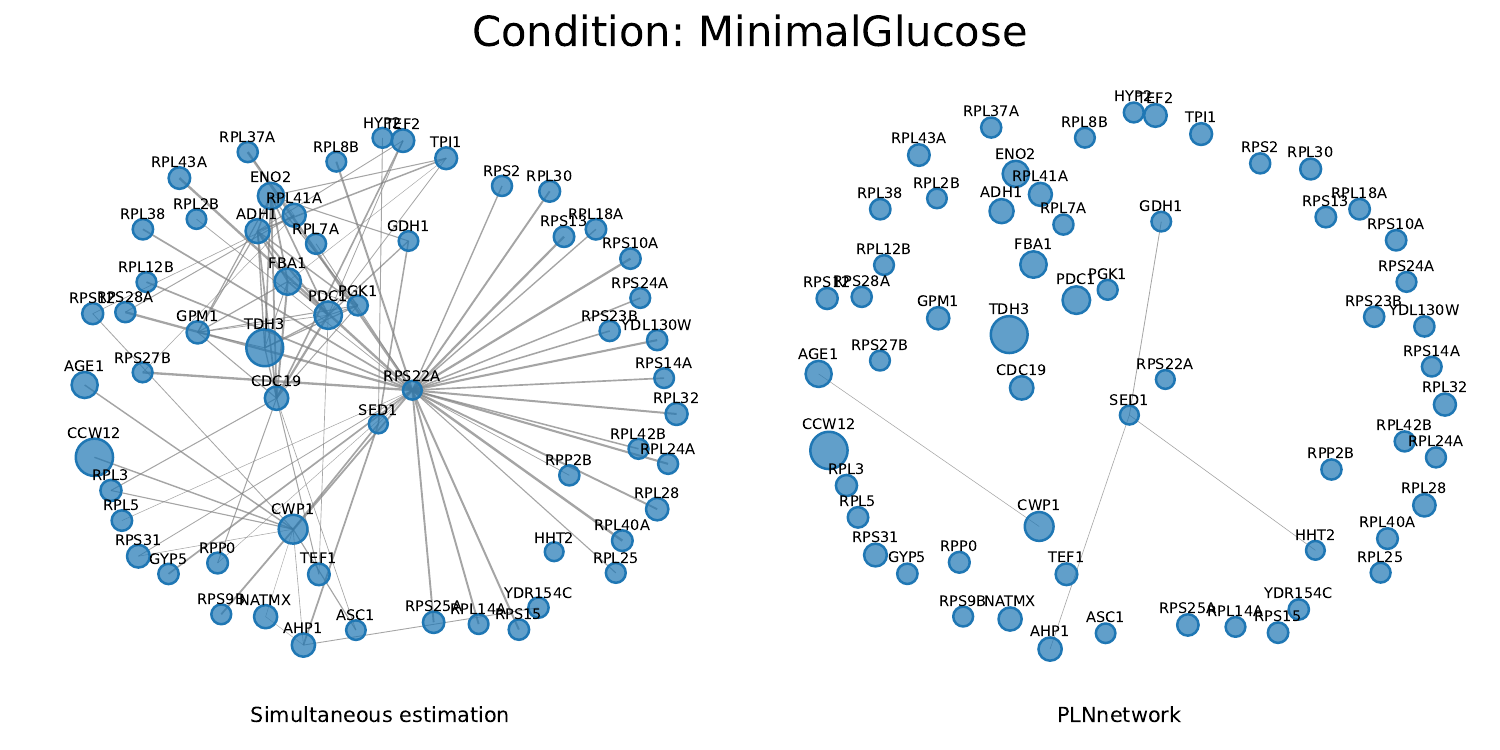}
\end{subfigure}
\begin{subfigure}[b]{.8\linewidth}
\includegraphics[width=\linewidth]{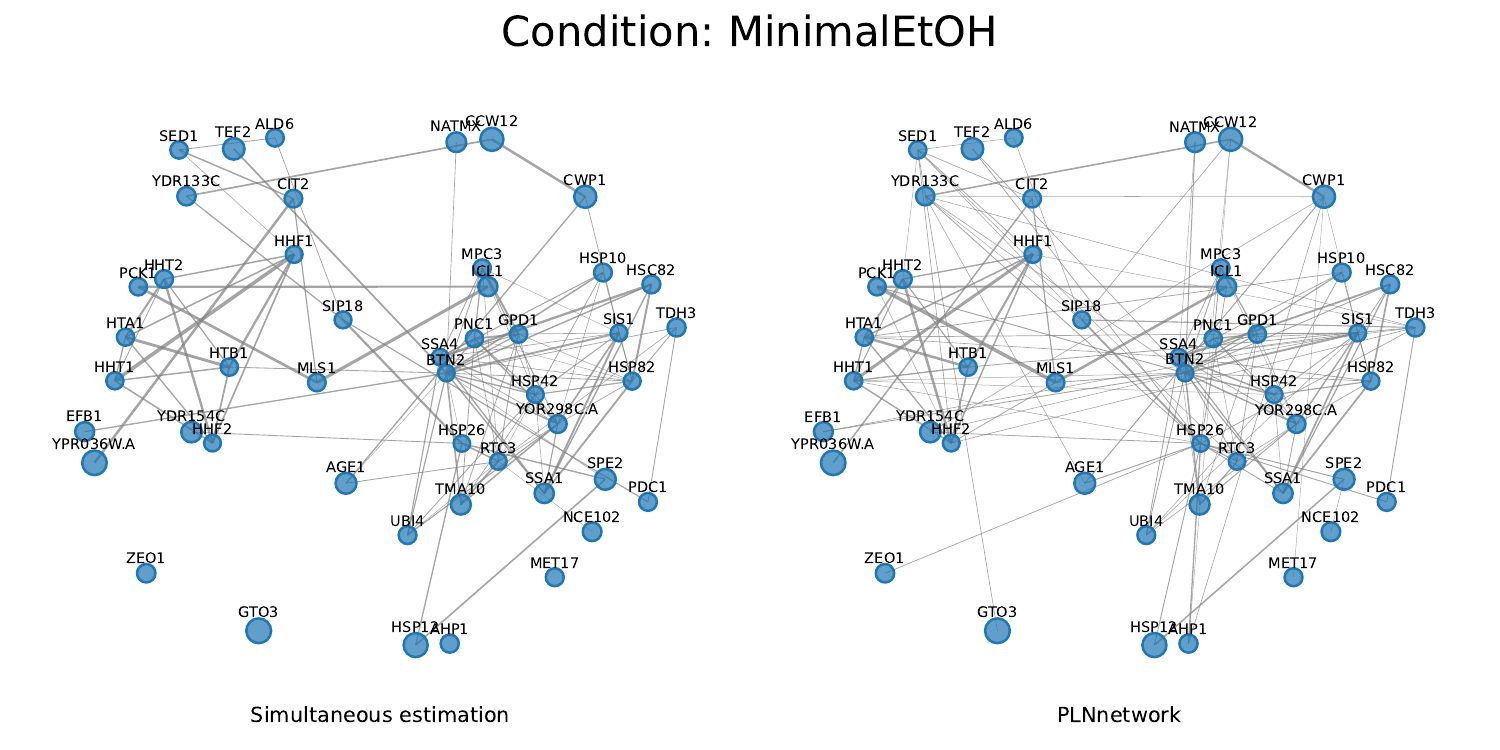}
\end{subfigure}
\begin{subfigure}[b]{.8\linewidth}
\includegraphics[width=\linewidth]{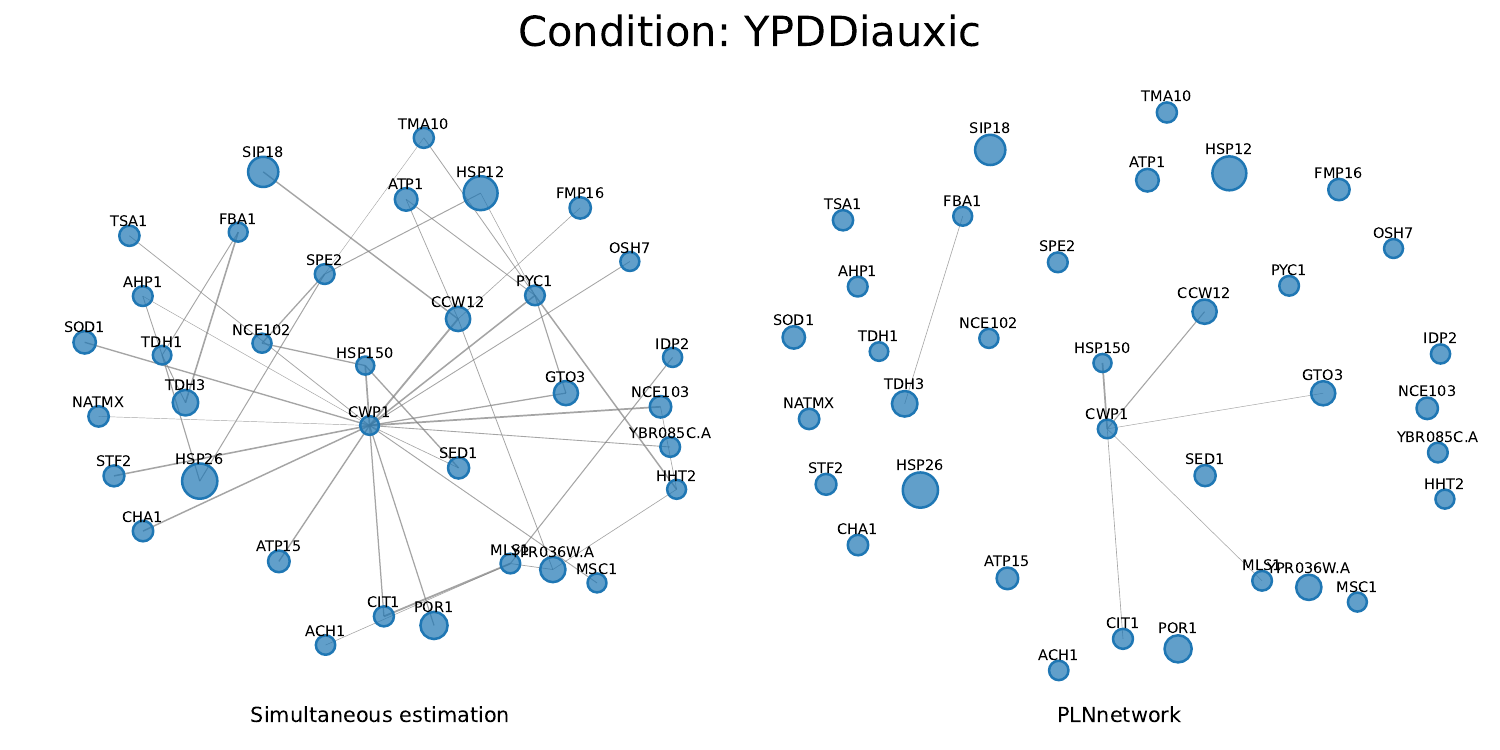}
\end{subfigure}
\caption{Gene expression network estimation for various conditions (cont.)}
\end{figure}

\clearpage
\begin{figure}
\centering
\ContinuedFloat
\begin{subfigure}[b]{.8\linewidth}
\includegraphics[width=\linewidth]{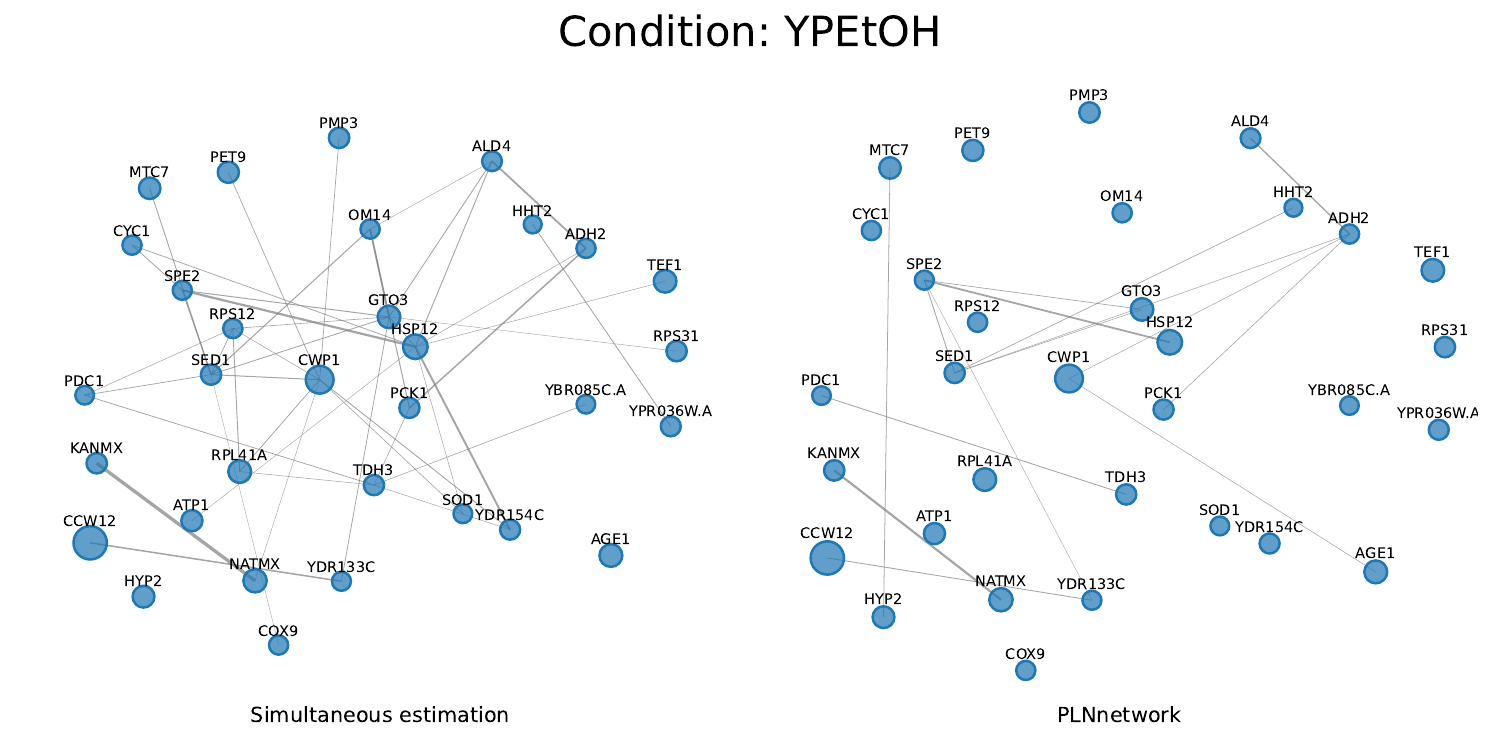}
\end{subfigure}
\caption{Gene expression network estimation for various conditions (cont.)}
\end{figure}

\begin{figure}
\centering
\begin{subfigure}[b]{.45\linewidth}
\includegraphics[width=\linewidth]{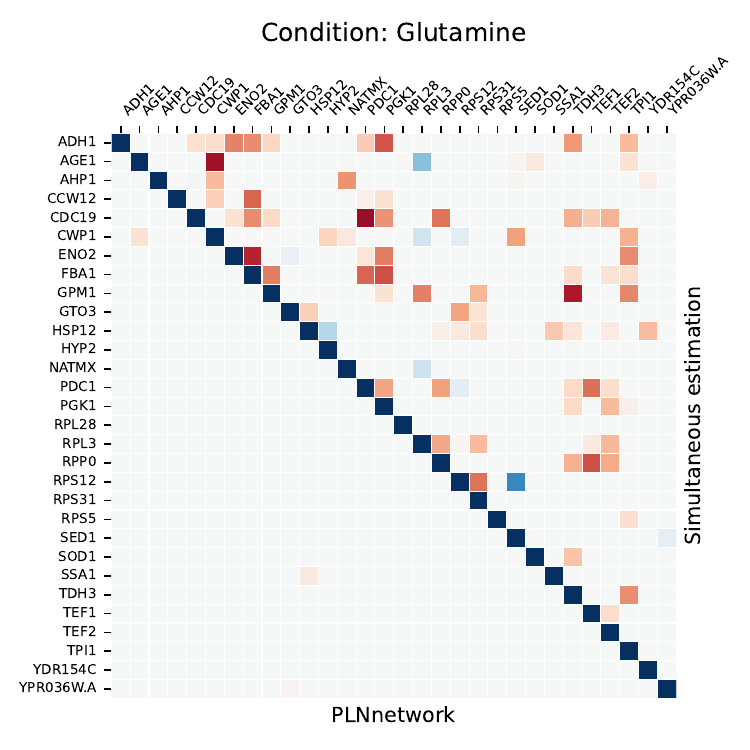}
\end{subfigure}
\begin{subfigure}[b]{.45\linewidth}
\includegraphics[width=\linewidth]{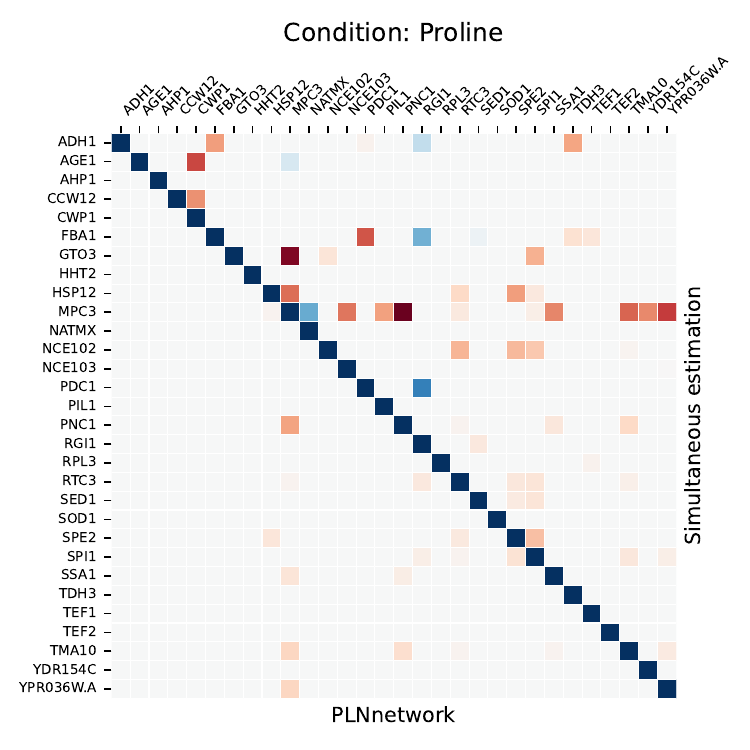}
\end{subfigure}
\caption{Conditional correlation matrices estimated by simultaneous estimation (upper right) and PLNnetwork (lower left) for various conditions. The legend is the same as Figure \plainref{fig:matrix_main}.}
\label{fig:matrix_app}
\end{figure}

\clearpage
\begin{figure}
\centering
\ContinuedFloat
\begin{subfigure}[b]{.45\linewidth}
\includegraphics[width=\linewidth]{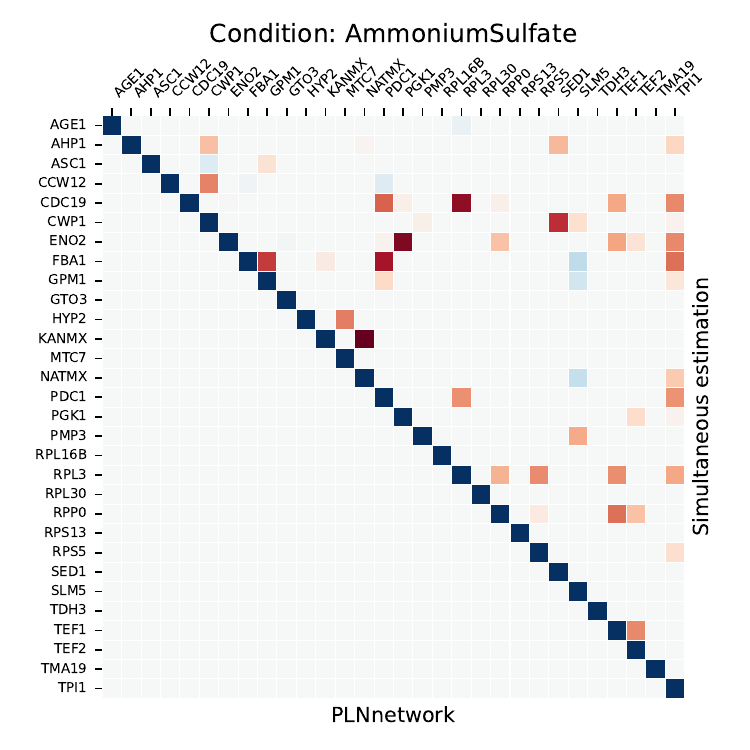}
\end{subfigure}
\begin{subfigure}[b]{.45\linewidth}
\includegraphics[width=\linewidth]{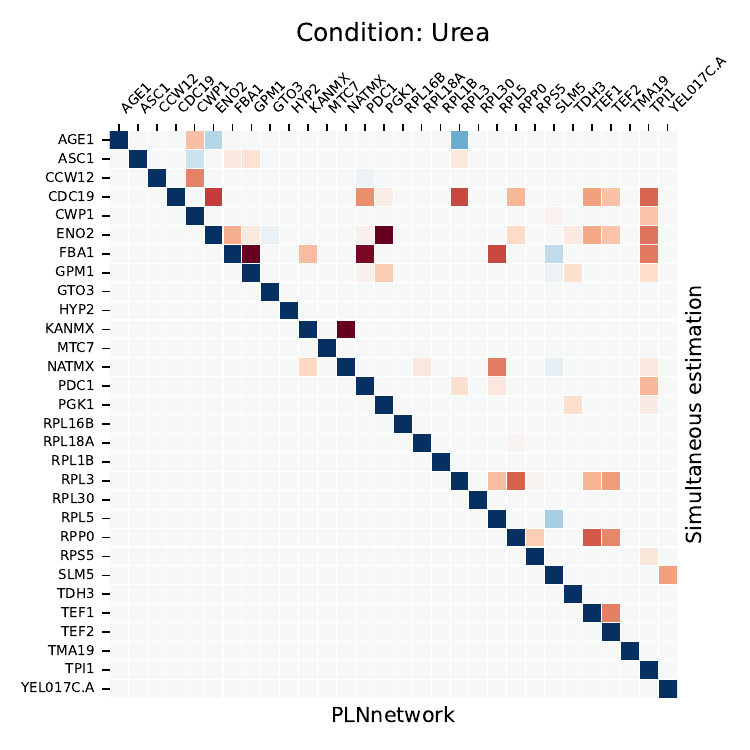}
\end{subfigure}
\begin{subfigure}[b]{.45\linewidth}
\includegraphics[width=\linewidth]{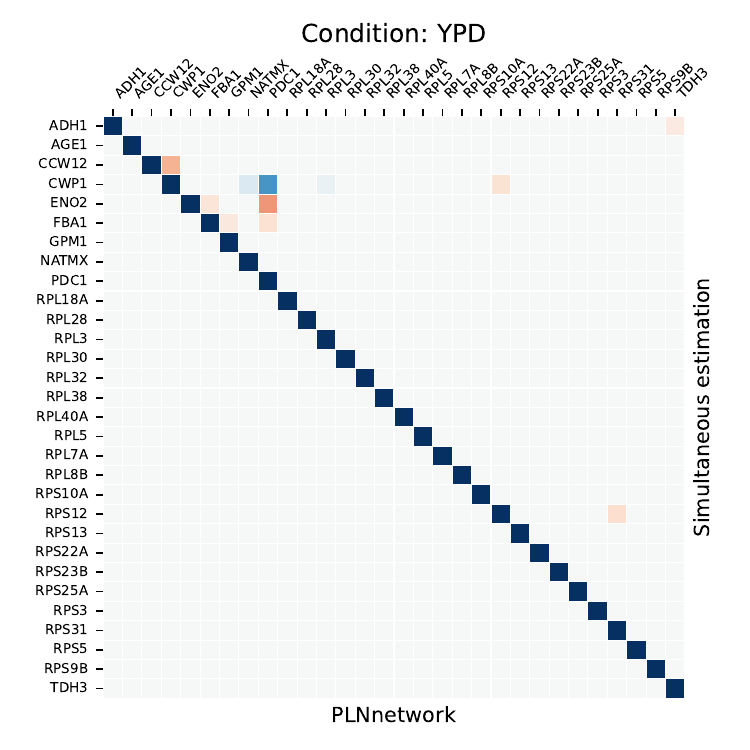}
\end{subfigure}
\begin{subfigure}[b]{.45\linewidth}
\includegraphics[width=\linewidth]{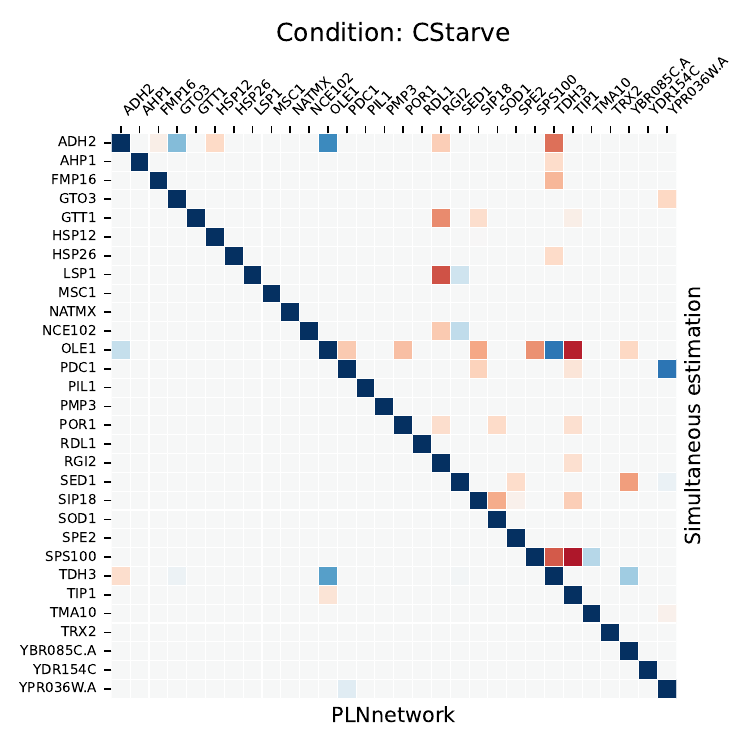}
\end{subfigure}
\caption{Conditional correlation matrices estimated by simultaneous estimation and PLNnetwork for various conditions (cont.)}
\end{figure}

\clearpage
\begin{figure}
\centering
\ContinuedFloat
\begin{subfigure}[b]{.45\linewidth}
\includegraphics[width=\linewidth]{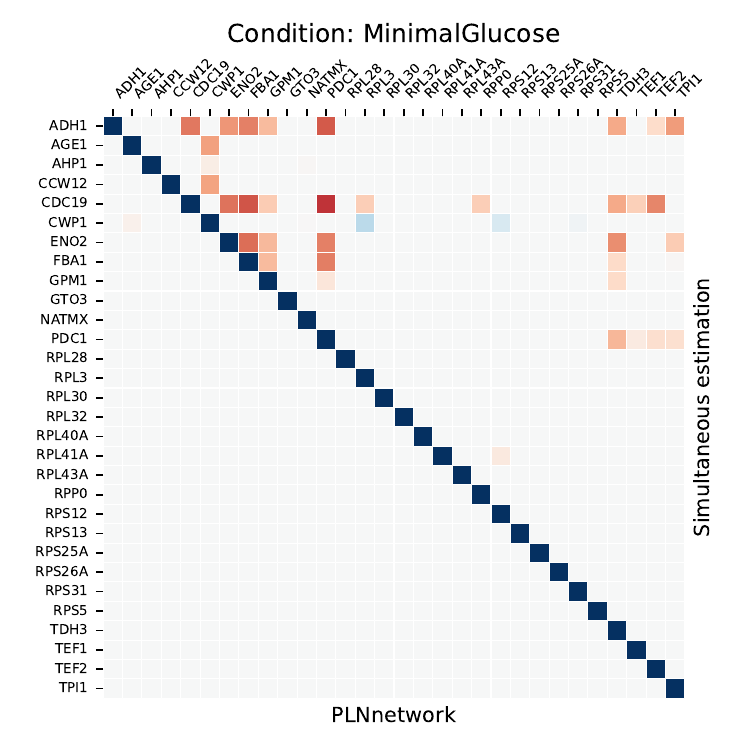}
\end{subfigure}
\begin{subfigure}[b]{.45\linewidth}
\includegraphics[width=\linewidth]{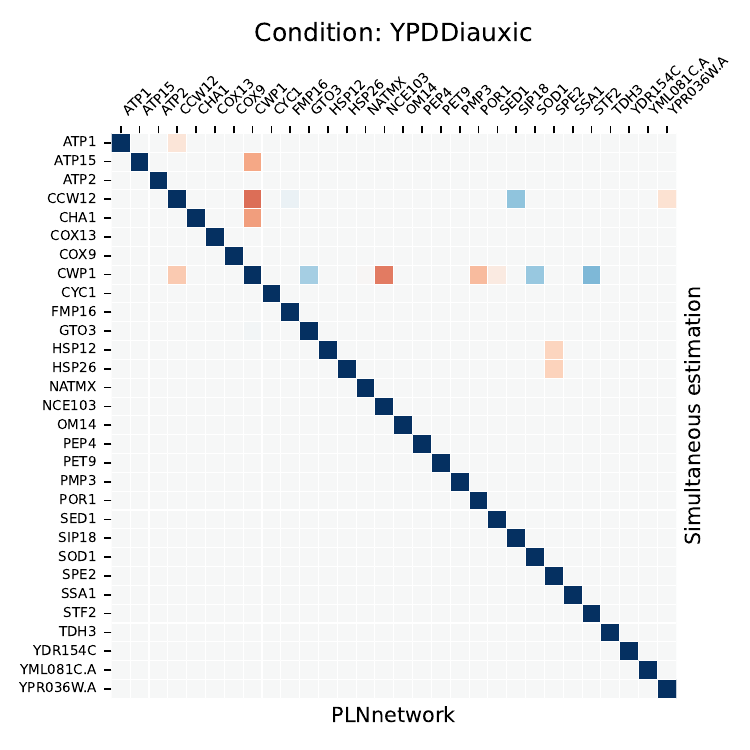}
\end{subfigure}
\begin{subfigure}[b]{.45\linewidth}
\includegraphics[width=\linewidth]{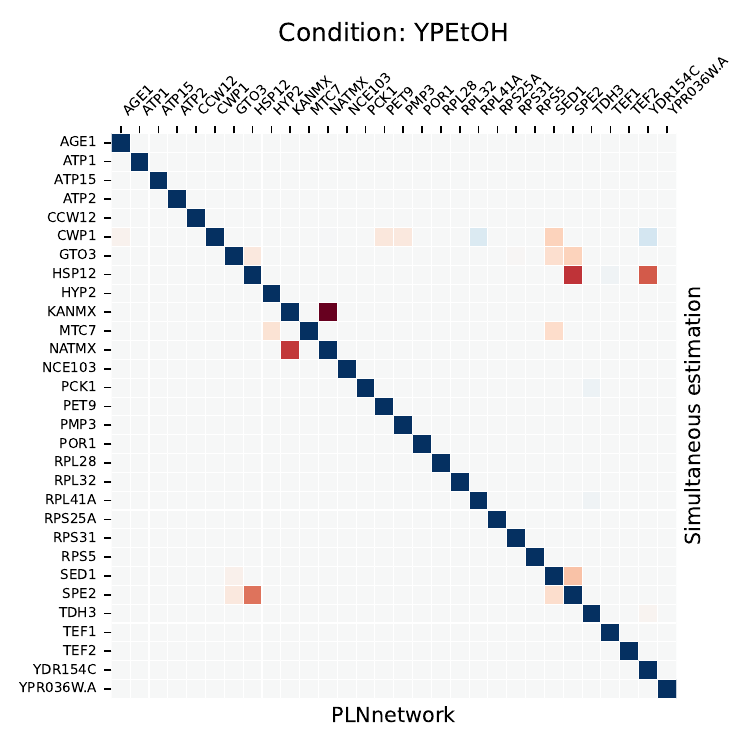}
\end{subfigure}
\begin{subfigure}[b]{.45\linewidth}
\includegraphics[width=\linewidth]{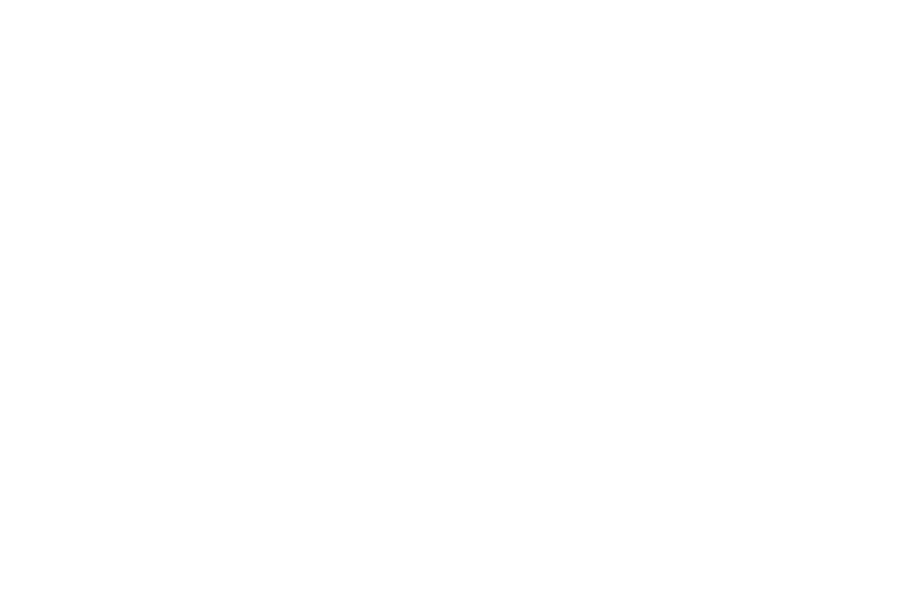}
\end{subfigure}
\caption{Conditional correlation matrices estimated by simultaneous estimation and PLNnetwork for various conditions (cont.)}
\end{figure}

\begin{figure}[h]
    \centering
    \includegraphics[width=.9\linewidth]{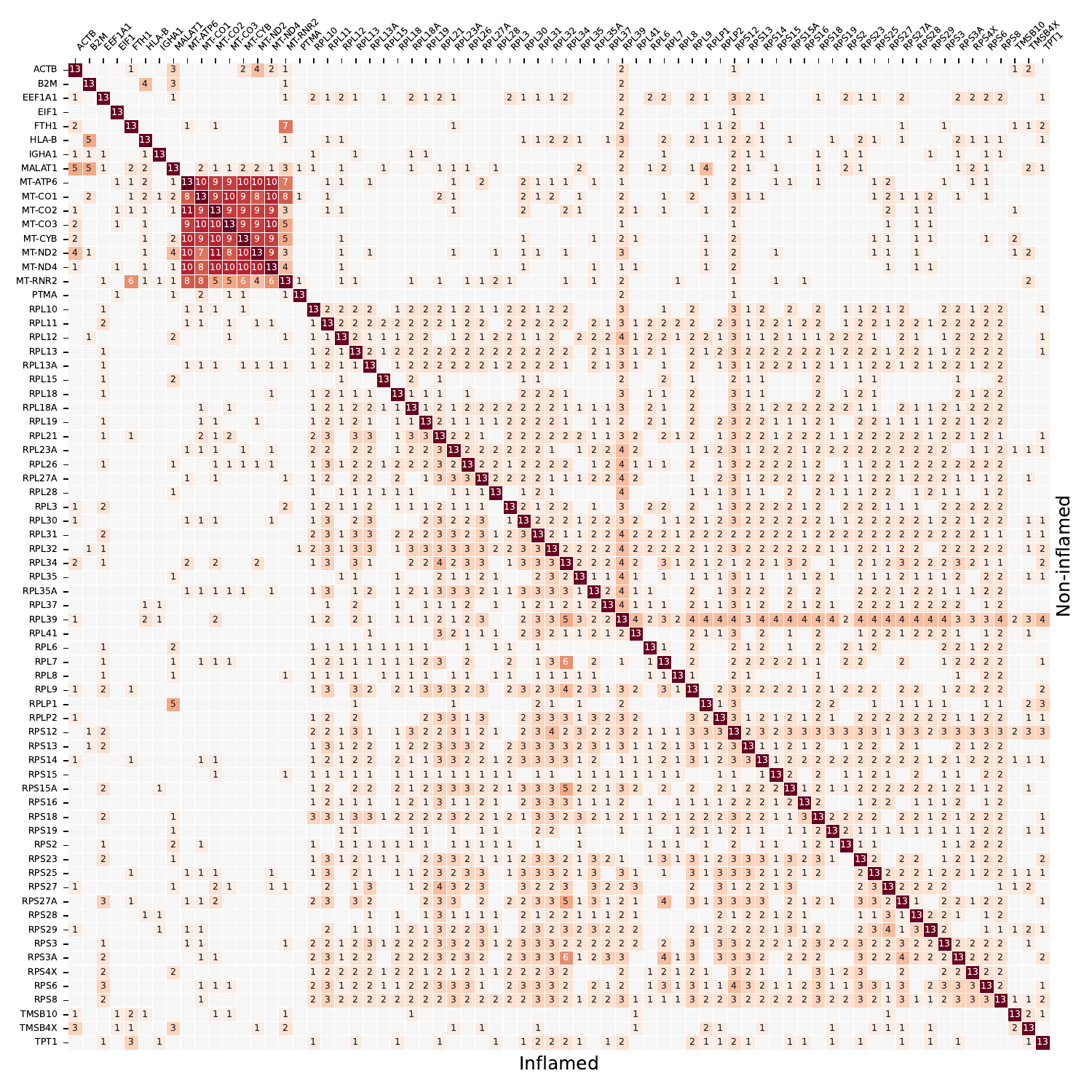}
    \caption{Comparison of gene expression networks in inflamed (lower left) vs. non-inflamed (upper right) cells of UC Patients by PLNnetwork. Each count indicates the number of patients, out of 13, having a link between each gene pair.}
    \label{fig:ibd_plnnetwork}
\end{figure}
\end{document}